\documentclass[aps,pra,twocolumn,superscriptaddress]{revtex4-2}

\usepackage{amsmath}
\usepackage{amssymb}
\usepackage{epsfig}
\usepackage{color}
\usepackage{subcaption}
\usepackage{adjustbox}
\usepackage{float}
\usepackage{verbatim}
\usepackage[bookmarks]{hyperref}  
\usepackage[normalem]{ulem}

\newcommand* {\ee}{\ensuremath{\mathrm{e}}}

\begin{document}

\title{Analytical evaluation of the coefficients of the Hu-Paz-Zhang master equation: Ohmic spectral density, zero temperature, and consistency check}

\author{G. Homa}
\email{homa.gabor@wigner.hu }
\affiliation{ Wigner Research Centre for Physics, Konkoly-Thege M.~\'ut 29-33, H-1121 Budapest, Hungary}
\affiliation{Department of Physics of Complex Systems, E\"{o}tv\"{o}s Lor\'and University, ELTE, P\'azm\'any P\'eter s\'et\'any 1/A, H-1117 Budapest, Hungary}
\author{J. Z. Bern\'ad}
\email{j.bernad@fz-juelich.de}
\affiliation{Peter Gr\"unberg Institute (PGI-8), Forschungszentrum J\"ulich, D-52425 J\"ulich, Germany}

\author{A. Csord\'as}
\email{csordas@tristan.elte.hu}
\affiliation{Department of Physics of Complex Systems, E\"{o}tv\"{o}s Lor\'and University, ELTE, P\'azm\'any P\'eter s\'et\'any 1/A, H-1117 Budapest, Hungary}

\date{\today}

\begin{abstract}
We investigate the exact master equation of Hu, Paz, and Zhang for a quantum harmonic oscillator at zero temperature with a Lorentz-Drude type Ohmic spectral density. This master equation plays an important role in the study of quantum Brownian motion and in various applications.
In this paper, we give an analytical evaluation of the coefficients of this non-Markovian master equation without Lindblad form, which allows us to investigate consistencies of the solutions, the positivity of the stationary density operator, and the boundaries of the model's parameters.  
\end{abstract}

\maketitle

\section{Introduction}
\label{I}
Quantum Brownian motion is a prototype example in the theory of open quantum systems \cite{Weiss, book1}.  The derivation of a master equation for the reduced density operator is central to the study of 
open quantum systems, and, in the case of the quantum Brownian motion \cite{Grabert} the first result was obtained by Caldeira and Leggett \cite{CL}.
Later an exact non-Lindblad master equation was derived by Hu, Paz, and Zhang (HPZ) with 
the help of path integral methods \cite{HuPazZhang92}. An alternative and simpler approach, which makes use of the Wigner function, was found by Halliwell and Yu \cite{Halliwell2012}, but see also Ref. \cite{Anglin}. All these methods yield a time-convolutionless master equation. 
A good overview of the history of the early results can be found in the book of Breuer and Petruccione \cite{book1}  or in the review article by Fleming, Roura, and Hu \cite{Fleming}.
The HPZ master equation can be subject to further approximations like the weak coupling or the high-temperature limit. In general, this 
time-convolutionless master equation is a special case of the Nakajima-Zwanzig equation, and perturbative expansions developed within this context can also be applied to get further master equations \cite{Nakajima, Zwanzig, Prigogine}. Therefore, the HPZ and its approximative master equations, including also the Caldeira-Leggett master equation, have a lot of applications in different branches of physics with a considerable literature. This includes for example the topic of magnets \cite{Anders}, spectroscopy \cite{Gottwald,Rognoni}, enzymes in a noisy environment \cite{Bothma}, quantum rate theory and dissipative tunneling \cite{Hanggi}, gravitationally induced decoherence \cite{Giesel, Jahn}, and finally, but not least, quantum field theory \cite{Yabu}.

The model under study consists of a central harmonic oscillator with mass $M$ and bare frequency $\Omega$ coupled to a thermal bath 
composed of further oscillators with mass $m_n$ and frequency $\omega_n$. The total Hamiltonian, which was already used in the early studies of dissipation \cite{FordKac, Ullersma, Ford01}, reads
\begin{equation}
 \hat{H}=\hat{H}_s+ \sum_n \hat{H}_n + \hat{H}_I, \nonumber
\end{equation}
where
\begin{eqnarray}
 \hat{H}_s&=& \frac{\hat{p}^2}{2M} +\frac12 M\Omega^2 \hat{q}^2, \nonumber \\
 \hat{H}_n&=& \frac{\hat{p}_n^2}{2m_n}+\frac12 m_n \omega_n^2 \hat{q}_n^2, \nonumber \\
 \hat{H}_I&=& \hat{q}\sum_n C_n \hat{q}_n,\label{eq:full_hamiltonian}
\end{eqnarray}
with $\hat{q},\hat{p}$ and $\hat{q}_n,\hat{p}_n$ being the coordinates and momenta of the central and bath oscillators, respectively. The central harmonic oscillator is coupled linearly to each bath 
oscillator with strength $C_n$.

All the information of the full system is contained in the density matrix $\rho(q,\mathbf{q},q',\mathbf{q}',t)\equiv \langle q,\mathbf{q}|\hat{\rho}(t)|q',\mathbf{q'}\rangle$, where we used the notation $\mathbf{q}=(q_1,q_2,\ldots)$. The equation of motion of the density operator $\hat{\rho}$ is given by the von Neumann equation
\begin{equation}
\frac{d}{dt}\hat{\rho}=-\frac{i}{\hbar}\left[\hat{H},\hat{\rho}\right]. 
\label{eq:neumann_eq}
\end{equation} 
The state of the central oscillator described by the reduced density matrix $\rho_s$ can be obtained by tracing out the degrees of freedom of the bath:
\begin{equation}
\rho_s(q,q',t)=\int \, d\mathbf{q} d\mathbf{q}' \, \delta(\mathbf{q}-\mathbf{q}')\rho(q,\mathbf{q},q',\mathbf{q}',t).
\label{eq:reduced_density_op}
\end{equation}
An equally good representation of the state of the central oscillator is given by the Wigner function defined as
\begin{equation}
W_s(q,p,t)=\frac{1}{2\pi\hbar}\int du \,e^{iup/\hbar} \rho_s(q-u/2,q+u/2,t).
\label{eq:def_of_wigner_function}
\end{equation}
The full system evolution is calculated under two basic assumptions: 

(i) The central and the bath oscillators are initially uncorrelated, which involves  the initial Wigner function 
 being factorized as

\begin{equation}
W(q,\mathbf{q},p,\mathbf{p},0)=W_s(q,p,0) \prod_n W_b(q_n,p_n,0).
\label{eq:factorization_of_wignerfunc} 
\end{equation}

(ii) The bath oscillators are initially in thermal equilibrium at temperature $T=1/(k_B \beta)$, which restricts the forms of $W_b$'s to 

\begin{equation}
W_b(q_n,p_n,0)=N_n\exp\left[-\frac{2}{\omega_n\hbar}\tanh\left(\frac{\hbar\omega_n\beta}{2}\right)H_n\right],
\label{eq:thermal_wigner_func}
\end{equation}
where $k_B$ is the Boltzmann-constant, $N_n$ is an appropriate normalization factor. 

The  Wigner function $W_s$ for the central  oscillator evolves according to a Fokker-Planck type equation,
\begin{eqnarray}
\frac{\partial W_s}{\partial t}&=& -\frac{p}{M} \frac{\partial W_s}{\partial q}+M\Omega^2 q\frac{\partial W_s}{\partial p}+A(t) q\frac{\partial W_s}{\partial p} \label{eq:master_equation_for_w} \\
&&+B(t)\frac{\partial (pW_s)}{\partial p}+C(t)\frac{\partial^2 W_s}{\partial p \partial q} +D(t) \frac{\partial^2 W_s}{\partial p^2},
\nonumber
\end{eqnarray}
which is equivalent to the HPZ master equation \cite{HuPazZhang92}. The time-dependent  real coefficients $A(t)$, $B(t)$, $C(t)$, and $D(t)$ strongly depend on the spectral density of the thermal bath 
\begin{equation}
I(\omega)=\sum_n \delta(\omega-\omega_n) \frac{C^2_n}{2 m_n \omega_n}.
\label{eq:def_of_spectral_density}
\end{equation}

The calculation of the coefficients in \eqref{eq:master_equation_for_w} is a rather complicated task and depends on solutions of the time-dependent equation of motion of the central oscillator. Exact formulas for the 
coefficients were obtained by Hu, Paz, and Zhang with the help of path integral methods in Ref. \cite{HuPazZhang92} and further simplified by Halliwell and Yu in Ref. \cite{Halliwell2012}. Simpler formulas have been derived from the beginning for the so-called weak coupling limit. As the central oscillator and the bath exchange energy in time, the oscillator magnitude is controlled by the couplings $C_n$. Thus, the 
weak coupling limit corresponds to the case when $C_n$'s are small. This limit can also be achieved as a systematic perturbation expansion of the time-convolutionless projection operator method \cite{Kappler}. These coefficients, being either exact or approximated, are used for the study of the stationary Gaussian state or to calculate averages of physical quantities and their standard deviations \cite{Fleming, Ford}. Most of the previous research performed a consistency check on the HPZ master equation by investigating only the stationary state, where the  Robertson-Schr\"odinger uncertainty principle, i.e., the positivity of the density operator, has to be fulfilled. In the case of the weak coupling limit with initial Gaussian states, we found that the analysis of the stationary state is not enough to get all the consistency conditions \cite{HCsCsB}. A full consistency check of the exact master equation is still missing, even though investigations or applications of the quantum Brownian motion have increased in the last decade; see for example Refs. \cite{Eisert, Paris, Lampo}. The mathematical reason to be careful is that this master equation is not in Lindblad form \cite{Lindblad} and even when its coefficients become constant the evolution is not described by a uniformly continuous semigroup \cite{Evans}. Therefore, this paper is devoted to the consistency study of solutions of the HPZ master equation at zero temperature with an Ohmic spectral density, which is considered to be a Lorentz-Drude type function with a high-frequency cutoff. We pick the $T=0$ case because this is the most interesting situation for describing decoherence and dissipation in a quantum Brownian motion \cite{Walls, Ford02}. 

In this paper, we evaluate analytically the exact coefficients $A(t)$, $B(t)$, $C(t)$, and $D(t)$.
The cornerstones of a rather long calculation are presented, where we also revisit the approach of Halliwell and Yu \cite{Halliwell2012}. One of the main hurdles is that we consider a more realistic spectral density than a purely Ohmic environment without a cutoff, which yields simple solutions of the generalized Langevin equation \cite{Ford}, but they lead to instantaneous dissipation, i.e., nonphysical behavior \cite{HuPazZhang92}. 

We study the stability and positivity of the asymptotic solutions of the master equation in three models with different couplings between the central system and the bath by including their behaviors in the weak coupling limit, too.
Furthermore, we will provide the short-time behaviors of the coefficients together with their asymptotics. It is known that all four time-dependent coefficients tend to a stationary value with time. When that happens the central harmonic oscillator evolves according to Markovian dynamics. Therefore, it is also worth knowing when the dynamics stop being non-Markovian and how this transition depends on the parameters of the model. 

The paper is organized as follows. In Sec. \ref{I.and.half} we discuss some inequalities from the point of view of stability and positivity of the solution of the master equation in the Markovian limit. In Sec. \ref{II}, we briefly recall the main results of Halliwell and Yu \cite{Halliwell2012}. This is followed up in Sec. \ref{III} by determining exactly the time-dependent coefficients, which are then studied
both analytically and numerically. In Sec. \ref{IV}, we study different models from the literature, which handle different couplings to the bath and the shift of the bare frequency of the central harmonic oscillator. Section \ref{V} summarizes the results and gives a concluding discussion. Long technical details are provided in four appendices.

\section{Consistency checks of the master equation's solutions in the Markovian $t \to \infty$ limit}
\label{I.and.half}

In this section, in the context of consistency check, we discuss the stability of the master equation in the Markovian limit and positivity of the steady state. Starting from a Gaussian $W_s$, the time-evolution given by Eq. \eqref{eq:master_equation_for_w} always keeps the Gaussian form for any time $t$. A further property of the HPZ master equation is that if the time evolution of the solution of Eq. (\ref{eq:master_equation_for_w}) is physical  under quite general conditions then there exists a characteristic time $\tau_M$ such that for $t \gg \tau_M$ the coefficients attain their asymptotic time-independent values, which is the so-called Markovian limit.
For shorter times the evolution is non-Markovian, i.e., the coefficients in \eqref{eq:master_equation_for_w} are time dependent. Let us analyze the final Markovian dynamics. It is best given in the representation
\begin{equation}
    \tilde{W}_s(k,\Delta,t)=\int_{-\infty}^{\infty}\int_{-\infty}^{\infty}\frac{dq\,dp}{(2\pi \hbar)^{1/2}} W_s(q,p,t)e^{i\left(k q+\frac{\Delta p}{\hbar}\right)},
\end{equation}
which is the so-called characteristic function of the central oscillator.
In the new representation, Eq. \eqref{eq:master_equation_for_w} reads as
\begin{eqnarray}
\frac{\partial \tilde{W}_s}{\partial t}&=& \Biggl(\frac{\hbar k}{M} \frac{\partial }{\partial\Delta}-\frac{M\Omega_\text{obs}^2}{\hbar} \Delta \frac{\partial }{\partial k} -2\lambda \, \Delta \frac{\partial }{\partial \Delta}\Biggr. \nonumber \\
&&\Biggl.-2D_{px}\, \Delta k-\frac{D_{pp}}{\hbar} \,\Delta^2  \Biggr)\tilde{W}_s,
\label{eq:eq_motion_for_W_k_delta}
\end{eqnarray}
with
\begin{eqnarray}
\frac{M\Omega_\text{obs}^2(\infty)}{\hbar}&=&\frac{M\Omega^2+A(\infty)}{\hbar},\quad 2\lambda=B(\infty),
\nonumber\\
2D_{p x}&=&\frac{C(\infty)}{\hbar},\quad \frac{D_{pp}}{\hbar}=\frac{D(\infty)}{\hbar^2}.
\label{eq:markovian_quantities}
\end{eqnarray}
Similar quantities, but at finite time $t$, [e.g., $2\lambda(t)=B(t)$, and so on] are called on page 477 in Ref.~\cite{book1}  physically observable frequency $\Omega_\text{obs}(t)$, damping coefficient  $\lambda(t)$, and diffusion coefficients $D_{px}(t),D_{pp}(t)$.
Let us use the notation $\mathbf{k}\equiv (\Delta,k)^T$. The initial condition is given as $\tilde{W}_s(\mathbf{k},t=0)=\tilde{W}_0(\mathbf{k})$. A correctly normalized $W_s(q,p,0)$ dictates that  $\tilde{W}_0$  at the origin $\mathbf{k}=\mathbf{0}$ should be equal to one, i.e., $\tilde{W}_0(\mathbf{0})=1$. Hence, Eq. \eqref{eq:eq_motion_for_W_k_delta} contains only first derivatives and thus can be solved with the method of characteristics \cite{Courant}. Here we quote the solution:
\begin{equation}
\tilde{W}_s(\mathbf{k},t)=\tilde{W}_0\Bigl(\exp{\left(-\mathbf{M}t\right)}\cdot\mathbf{k}\Bigr) \exp\left(-\mathbf{k}^T\cdot\tilde{\mathbf{R}}(t)\cdot\mathbf{k}\right).
\label{eq:markovian_solution}
\end{equation}
The matrix $\mathbf{M}$ can be expressed with its orthogonal projections $\mathbf{P}_i$, $\mathbf{P}_i \mathbf{P}_j =\delta_{i,j}\mathbf{P}_i$, $i=1,2$ as
\begin{equation}
\mathbf{M}=\begin{pmatrix}
2\lambda & -\frac{\hbar}{M} \\
\frac{M\Omega^2_\text{obs}(\infty)}{\hbar} & 0
\end{pmatrix}=\Lambda_1 \mathbf{P}_1+\Lambda_2 \mathbf{P}_2,
\end{equation}
where the eigenvalues $\Lambda_i$ are
\begin{equation}
\Lambda_{1,2}=\lambda \pm \sqrt{\lambda^2-\Omega^2_\text{obs}(\infty)}, 
\label{eq:markovian_eigenvalues}
\end{equation}
and the projections are given by
\begin{equation}
\mathbf{P}_i=\frac{\Omega^2_\text{obs}(\infty)}{\Omega^2_\text{obs}(\infty)-\Lambda_i^2} \begin{pmatrix}
-\frac{\Lambda_i^2}{\Omega^2_\text{obs}(\infty)} & \frac{\hbar\Lambda_i}{\Omega^2_\text{obs}(\infty) M} \\ 
\\
\frac{M\Lambda_i}{\hbar} &  1
\end{pmatrix}, \quad i=1,2.
\end{equation}
The first factor in \eqref{eq:markovian_solution} ensures that the stationary distribution $\tilde{W}_s(\mathbf{k},t=\infty)$ is independent of the initial condition $\tilde{W}_0(\mathbf{k})$ 
chosen if the limiting matrix is zero: $\lim_{t \to \infty}\exp{\left(-\mathbf{M}t\right)}=0$. 
In that case due to $\tilde{W}_0(\mathbf{0})=1$ the first factor in \eqref{eq:eq_motion_for_W_k_delta} is asymptotically $1$. This requires that real parts of the two eigenvalues $\Lambda_i$ should be non-negative, otherwise the time evolution of the solution $\tilde{W}_s(\mathbf{k},t)$ will not converge to a unique asymptotic characteristic function $\tilde{W}_s(\mathbf{k},\infty)$. From  \eqref{eq:markovian_eigenvalues} it follows that the following inequalities must hold:
\begin{equation}
\lambda\geq 0, \qquad \Omega^2_\text{obs}(\infty) \geq 0,
\label{eq:om_obs2_pos}
\end{equation}
which we call the stability conditions. Straightforward calculations lead to the matrix $\tilde{\mathbf{R}}(t)$ in the second factor of \eqref{eq:markovian_solution} to
\begin{eqnarray}
    \tilde{\mathbf{R}}(t)&=& \left( \mathbf{P}_1^T \mathbf{R} \mathbf{P}_2+\mathbf{P}_2^T \mathbf{R} \mathbf{P}_1\right) \frac{1-e^{-(\Lambda_1+\Lambda_2)t}}{\Lambda_1+\Lambda_2} \nonumber \\
    &&+\sum_{i=1}^2 \mathbf{P}_i^T \mathbf{R} \mathbf{P}_i \frac{1-e^{-2\Lambda_i}}{2\Lambda_i},
\end{eqnarray}
where the constant matrix $\mathbf{R}$ is built up from the diffusion coefficients as
\begin{equation}
    \mathbf{R}=\begin{pmatrix} 
        \frac{D_{pp}}{\hbar} & D_{p x} \\
        D_{p x} & 0
    \end{pmatrix}.
\end{equation}
Using the above formulas the asymptotic matrix 
\begin{equation}
\tilde{\mathbf{R}}(\infty)=\lim_{t \to \infty} \tilde{\mathbf{R}}(t)=\begin{pmatrix} 
        \frac{D_{pp}}{4\hbar\lambda} & 0 \\
        0 & \frac{\hbar\left( D_{p p}+4 D_{p x} \lambda M\right)}{4\lambda M^2 \Omega_\text{obs}^2(\infty)}
    \end{pmatrix}.
\end{equation}
must be positive, otherwise $\tilde{W}_s(\mathbf{k},\infty)$ tends to infinity as $|\mathbf{k} |\to \infty$. This requirement implies that
\begin{equation}
\frac{D_{pp}}{4\hbar\lambda}\geq 0, \quad \text{and} \quad \frac{\hbar\left( D_{p p}+4 D_{p x} \lambda M\right)}{4\lambda M^2 \Omega_\text{obs}^2(\infty)} \geq 0.
\label{eq:weak_condition}
\end{equation}
Conditions in Eqs. \eqref{eq:om_obs2_pos} and \eqref{eq:weak_condition} are necessary so that the solution of the
master equation \eqref{eq:master_equation_for_w} in the limit $t \to \infty$ tends to a stationary Gaussian function decaying to zero as $|\mathbf{k}| \to \infty$. However, this asymptotic Gaussian solution does not necessarily describe a physical situation belonging to a density operator. 


The eigenvalue problem in coordinate representation of a  bivariate, self-adjoint  and trace-class  operator $\rho(x,y)=\langle x |\hat \rho |y\rangle $ is 
\begin{equation}
\int_{-\infty}^{\infty} \rho(x,y) \phi_n(y) dy =\lambda_n \phi_n(x).
\label{eq:eigenvalue_problem}
\end{equation}
For Gaussian $\rho(x,y)$ with unit trace the eigenvalues and eigenvectors were determined in 
 Refs.~\cite{HCSB,Newton}. The criterion there that all the eigenvalues are in the interval $[0,1]$, which is called positivity criterion for a density matrix, can be stated for $\tilde{W}_s(\mathbf{k},t=\infty)$ as follows: the requirement  $16\,\textrm{Det}(\tilde{\mathbf{R}}(\infty)) \geq 1$ should be also true, which reads as 
\begin{equation}
Q\equiv \frac{4D(\infty)\Bigl( D(\infty)+M C(\infty)B(\infty)\Bigr)}{\hbar^2 M^2 B^2(\infty)\Omega_\text{obs}^2(\infty)}\geq 1.
\label{eq:Q_def}
\end{equation}

A little calculation shows, that this inequality for the asymptotic Gaussian state is the same as that used as a complete positivity condition in bosonic Gaussian channels in Eqs. (7) and (9) of Ref. \cite{caruso2018}, namely $\sigma_y+4\tilde{R}(\infty) \geq 0$, where $\sigma_y$ is the second Pauli matrix.

In the case of non-Gaussian initial states, the positivity check is a difficult problem, because the analytical solution of the eigenvalue equation \eqref{eq:eigenvalue_problem} is generally not known. However, positivity can be monitored by the method of Ref.~\cite{Newton}, which uses the different moments of $\hat{\rho}$ and Newton's identities by checking the positivity of infinitely many scalar quantities. This method can always be applied for the solution of the master equation at any time starting from a non-Gaussian initial $\rho(x,y,t=0)$.   


\section{The forms of the time-dependent coefficients}
\label{II}

In this section, we summarize the results of Halliwell and Yu \cite{Halliwell2012}. This approach leads indeed to exact solutions of the coefficients compared to the
attempt based on a local in-time approximation of the central oscillator's equation of motion \cite{Ford}. The time-dependent coefficients $A(t)$, $B(t)$, $C(t)$, and $D(t)$ are determined by deriving and solving the differential equations for the central oscillator as
\begin{equation}
\frac{d^2}{ds^2}{q}(s)+\Omega^2 q(s)+\frac{2}{M} \int_0^s d \lambda \,\eta(s-\lambda) q(\lambda)=\frac{f(s)}{M}
\label{eq:diff_eqs_for_q(t)}
\end{equation}
with appropriate initial conditions. The form of $f(s)$ is fixed by the statistical properties of the bath of oscillators. The exact expressions for the  coefficients $A(t)$, $B(t)$, $C(t)$, and $D(t)$, including all the orders of the interactions with the bath, requires the solution of \eqref{eq:diff_eqs_for_q(t)} in two cases: (i) the forward solution for $s>0$ when $q(0)$ and $\dot{q}(0)=p(0)/M$ are fixed and (ii) the backward solution for $s<t$ when $q(t)$ and $\dot{q}(t)=p(t)/M$ are fixed as initial conditions. 

In Eq. \eqref{eq:diff_eqs_for_q(t)} the temperature-independent kernel $\eta(s)$ is fixed by spectral density as
\begin{equation}
\eta(s)=-\int_0^\infty d\omega\, I(\omega)\sin(\omega s).
\label{eq:def_eta}
\end{equation}
$C(t)$ and $D(t)$ bring a new  temperature-dependent kernel $\nu(s)$ into the theory, which is
\begin{equation}
\nu(s)
=\int_0^\infty d\omega\, I(\omega)\coth\left(\frac{\hbar\omega\beta}{2}\right)\cos(\omega s).
\label{eq:def_nu}
\end{equation}

A few comments are in order. At the heart of the present problem is how to solve \eqref{eq:diff_eqs_for_q(t)} with appropriate boundary conditions. In Refs. \cite{Ford,book1,Fleming} an equivalent form of Eq. \eqref{eq:diff_eqs_for_q(t)} was used by performing integration by parts in the convolution of $\eta(s)$ and $q(s)$. In that way, a so-called generalized Langevin equation ought to be solved, and several initial conditions have been proposed to solve it. However, exact solutions must be equal in all versions of the formalisms. To our best knowledge, the solution of Halliwell and Yu is exact, and by using the backward solutions they were able to give a full account of the interplay between the central oscillator and the bath. 

The coefficients $A(t)$, $B(t)$, $C(t)$, and $D(t)$ of the master equation can be determined by the time evolution of the first and second moments of $p$ and $q$. For the details see Refs. \cite{Halliwell2012,Ford}. Then, we have that $A(t)$ and $B(t)$ are temperature independent and they are given by \cite{Halliwell2012}
\begin{equation}
A(t)=2\int_0^t ds\, \eta(t-s)u_2(s)-2\frac{\dot{u}_2(t)}{\dot{u}_1(t)} \int_0^t ds\, \eta(t-s)u_1(s)
\label{eq:def_of_a(t)}
\end{equation}
and
\begin{equation}
B(t)=\frac{2}{M\dot{u}_1(t)} \int_0^t ds\, \eta(t-s)u_1(s),
\label{eq:def_of_b(t)}
\end{equation}
respectively. Here in Eqs. \eqref{eq:def_of_a(t)} and \eqref{eq:def_of_b(t)} the functions $u_1(t)$ and $u_2(t)$ are two elementary functions, which are the two solutions of the homogeneous, linear integrodifferential equation corresponding to
\eqref{eq:diff_eqs_for_q(t)},
\begin{equation}
\frac{d^2}{ds^2}{u}(s)+\Omega^2 u(s)+\frac{2}{M} \int_0^s d \lambda \,\eta(s-\lambda) u(\lambda)=0.
\label{eq:diff_eqs_for_u}
\end{equation}
Boundary conditions for $u_1$ and $u_2$ are
\begin{eqnarray}
u_1(s=0)&=&1, \quad u_1(s=t)=0, \label{eq:bound_cond_for_u1} \\
u_2(s=0)&=&0, \quad u_2(s=t)=1. \label{eq:bound_cond_for_u2}
\end{eqnarray}
The coefficients $C(t)$ and $D(t)$ depend on temperature and require also the knowledge of two Green's functions, which are the solutions of the inhomogeneous, linear integrodifferential equation connected to \eqref{eq:diff_eqs_for_q(t)},
\begin{eqnarray}
&&\frac{d^2}{ds^2}{G_i}(s,\tau)+\Omega^2 G_i(s,\tau)+\frac{2}{M} \int_0^s d \lambda \,\eta(s-\lambda) G_i(\lambda,\tau) \nonumber \\
&&\qquad \qquad=\delta(s-\tau), \quad i=1,2.
\label{eq:diff_eqs_for_G}
\end{eqnarray}

Boundary conditions for $G_1$ and $G_2$ are prescribed by
\begin{eqnarray}
G_1(s=0,\tau)&=&0, \quad \frac{d}{ds}G_1(s=0,\tau)=0, \label{eq:bound_cond_for_G1} \\
G_2(s=t,\tau)&=&0, \quad \frac{d}{ds}G_2(s=t,\tau)=0. \label{eq:bound_cond_for_G2}
\end{eqnarray}
It is worth mentioning that the boundary conditions in \eqref{eq:bound_cond_for_G2} are not explicitly shown by Ref. \cite{Halliwell2012}. 

Thus, the coefficients $C(t)$ and $D(t)$ are
\begin{widetext}
\begin{equation}
C(t)=\frac{\hbar}{M}\int_0^\infty d\lambda\, G_1(t,\lambda)\nu(t-\lambda)-\frac{2\hbar}{M^2} \int_0^t ds \int_0^\infty d\tau \int_0^\infty d\lambda \, \eta(t-s)G_1(t,\lambda)G_2(s,\tau)\nu(\tau-\lambda), 
\label{eq:def_of_c(t)}
\end{equation}
\begin{equation}
D(t)=\hbar\int_0^\infty d\lambda\, G'_1(t,\lambda)\nu(t-\lambda)-\frac{2\hbar}{M} \int_0^t ds \int_0^\infty d\tau \int_0^\infty d\lambda \, \eta(t-s)G'_1(t,\lambda)G_2(s,\tau)\nu(\tau-\lambda),
\label{eq:def_of_d(t)}
\end{equation}
\end{widetext}
where the signs of the triple integrals are minus compared to \cite{Halliwell2012}, but they agree with the preprint version \cite{Halliwellarxiv}.
The prime on $G_1(t,\lambda)$ means derivative with respect to the first variable of $G_1$. Equations \eqref{eq:def_of_a(t)}, \eqref{eq:def_of_b(t)}, \eqref{eq:def_of_c(t)} and 
\eqref{eq:def_of_d(t)} give the exact coefficients of the master equation in our problem.

An important special case, which is widely studied in the literature (c.f. \cite{book1}), is the case of weak coupling, i.e., the coupling constants $C_n$ in \eqref{eq:full_hamiltonian} are small.
By calculating the coefficients in leading order in the coupling constant $C_n$, Halliwell and Yu also obtained the formulas of a consistent weak coupling limit. 
Now, the time-dependent coefficients read as
\begin{eqnarray}
A_w(t)&=&2\int_0^t ds \, \eta(s) \cos(\Omega s),
\label{eq:a(t)_in_weak_coupling} \\
B_w(t)&=&-\frac{2}{M\Omega}\int_0^t ds \, \eta(s) \sin(\Omega s),
\label{eq:b(t)_in_weak_coupling} \\
C_w(t)&=&\frac{\hbar}{M\Omega}\int_0^t ds \, \nu(s) \sin(\Omega s),
\label{eq:c(t)_in_weak_coupling} \\
D_w(t)&=&\hbar\int_0^t ds \, \nu(s) \cos(\Omega s).
\label{eq:d(t)_in_weak_coupling}
\end{eqnarray}    
where we have indicated this approximation with the index $w$. We note that the derivation of the weak coupling limit requires the expansion of $q(t)$ and $q_n(t)$ up to the second order of $C_n$. In the next section, we make use of these compact formulas and calculate explicitly the exact coefficients. 

\section{Exact solutions to $\textit{\textbf{A(t)}}$, $\textit{\textbf{B(t)}}$,  $\textit{\textbf{C(t)}}$, and $\textit{\textbf{D(t)}}$}
\label{III}

\subsection{Calculation of $u_1$, $u_2$, $G_1$, $G_2$}
\label{subsec:cal_of_u1_u2_g1_g2}

To solve the integrodifferential equations  \eqref{eq:diff_eqs_for_u} and \eqref{eq:diff_eqs_for_G}, one needs to pick a spectral density. Here, we consider an Ohmic spectral density with a Lorentz-Drude type function \cite{Weiss,book1,HCsCsB}
and a high-frequency cutoff $\Omega_c$:
\begin{equation}
I(\omega)=\frac{2M\gamma \Omega_c^2}{\pi} \frac{\omega}{\omega^2+\Omega_c^2},
\label{eq:Ohmic_spectral_density}
\end{equation}
where $\gamma$ is a frequency-independent constant. The cutoff frequency $\Omega_c$ is large compared to the bare oscillator frequency $\Omega$. An increase in the couplings $C_n$ implies a bigger value for $\gamma$. We note that some authors call $\gamma$ a damping constant because at large temperature $T$ the effective damping constant (the damping constant for a long time $t$) is almost equal to $\gamma$ (see, e.g., \cite{book1}). 

The temperature-independent kernel in Eq. \eqref{eq:def_eta} is an odd function of $s$. For $s>0$ it reads
\begin{equation}
\eta(s)=-M \gamma \Omega_c^2 e^{-\Omega_c s}, \qquad s>0.
\label{eq:result_for_eta}
\end{equation}
The other kernel $\nu(s)$ in Eq. \eqref{eq:def_nu} is even as a function of $s$. It is evaluated at a finite temperature usually by expanding the hyperbolic cotangent function and integrating term by term for $\omega$. Here we are interested in the zero-temperaturelimit, which simply means that 
the hyperbolic cotangent function in \eqref{eq:def_nu} is replaced by 1:
\begin{equation}
\nu(s)
=\int_0^\infty d\omega\, I(\omega)\cos(\omega s), \qquad T=0.
\label{eq:def_nu_at_T_eq_zero}
\end{equation}
It can be given explicitly; however, our strategy is to insert \eqref{eq:def_nu_at_T_eq_zero} into the expressions of the coefficients of the master equation and perform the $\omega$ integral as the last step. 

Now, as we have established the function $\eta(s)$, Eq. \eqref{eq:diff_eqs_for_u} can be solved with the
help of the Laplace transformation 
\begin{equation}
\tilde{u}(z)= {\cal L}_s\left[u(s)\right](z)\equiv \int_0^\infty u(s) e^{-sz}\, ds.
\label{eq:laplace_trafo}
\end{equation}
This leads to
\begin{equation}
\left( \Omega^2 +z^2 -\frac{2\gamma \Omega_c^2}{\Omega_c+z}\right)\tilde{u}(z)-u'(0)-z u(0)=0. 
\label{eq:u_after_laplace}
\end{equation}
Values for $u(0)$ and $u'(0)$ will be fixed after performing the inverse Laplace transformation
\begin{equation}
u (s)= {\cal L}_z^{-1}\left[\tilde{u}(z)\right](s)\equiv \frac{1}{2 \pi i} 
\int_{\mathcal{C}} \tilde{u}(z) \ee^{zs}\, dz.  
\label{eq:inverse_laplace_trafo}
\end{equation}
The path of integration $\mathcal{C}$ has to 
be chosen in such a way that all poles of $\tilde{u}(z)$
are included.

Let us write the combination in parentheses of Eq. \eqref{eq:u_after_laplace} as
\begin{equation}
\Omega^2 +z^2 -\frac{2\gamma \Omega_c^2}{\Omega_c+z} \equiv \frac{(z-z_1)(z-z_2)(z-z_3)}{\Omega_c+z},
\label{eq:replacement_in_lapl_transform}
\end{equation}
where $z_1,z_2,z_3$ are the three roots of the cubic equation
\begin{equation}
z^3+\Omega_c z^2+\Omega^2z+\Omega^2\Omega_c-2\gamma\Omega_c^2=0.
\label{eq:cubic_equation}
\end{equation}
\begin{figure}
	\includegraphics[bb= -100pt -50pt 554pt 770pt,angle=270,width=1.0 \hsize]{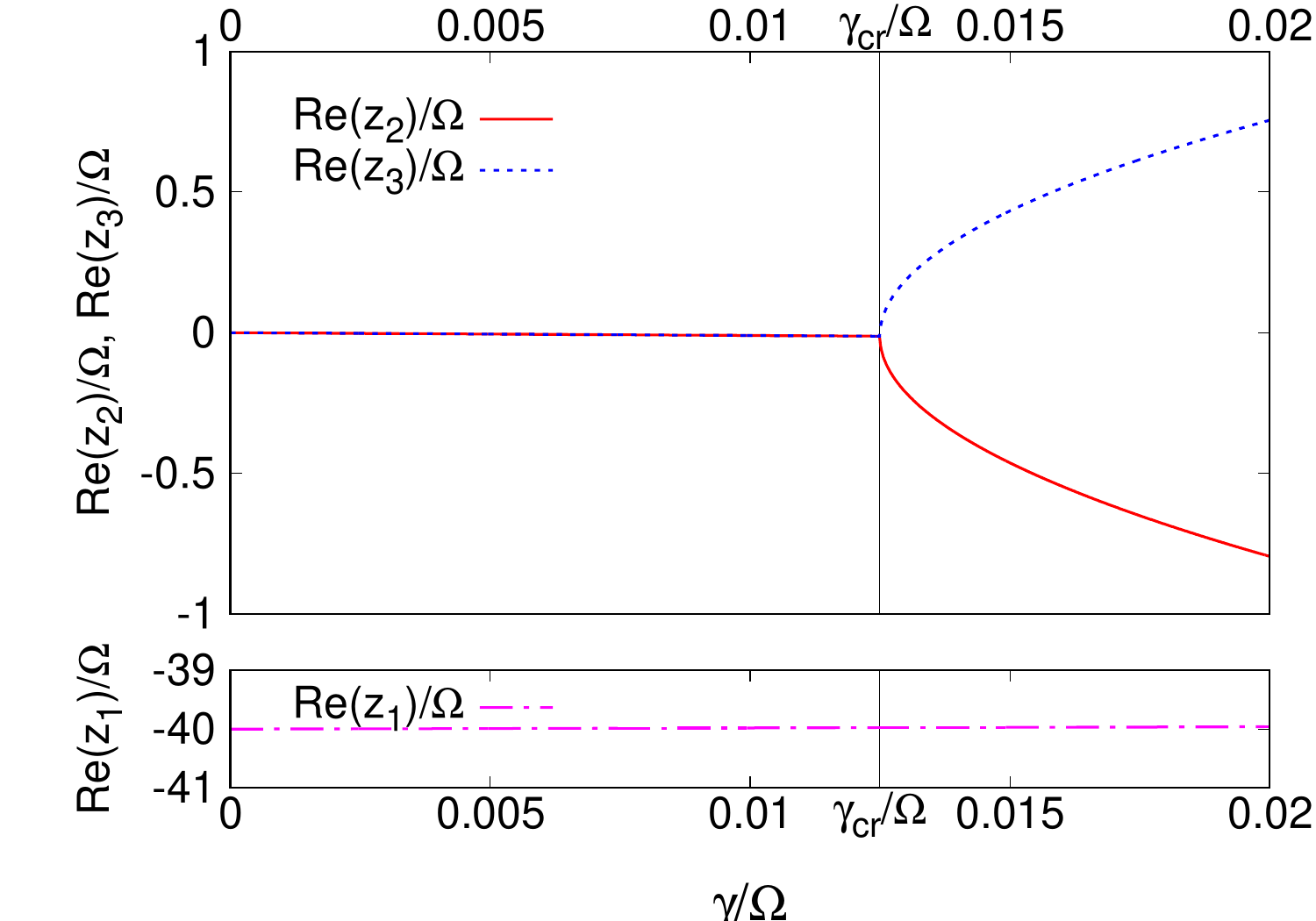}
	\caption{ Real parts of the three roots of Eq. (\ref{eq:cubic_equation}) as a function of 
	dimensionless coupling constant $\gamma/\Omega$. For both figures we have set the cutoff frequency $\Omega_c=40\Omega$.
	 The vertical line indicates the critical coupling constant $\gamma_{\text{cr}}$ in \eqref{eq:gamma_crit_first_model}.\label{fig:roots_of_orig_model}}
\end{figure}
In Fig.~\ref{fig:roots_of_orig_model} we show a typical ($\Omega_c \gg \Omega$) plot for the real parts of the roots. At 
\begin{equation}
\gamma_{\text{cr}}=\Omega^2/(2\Omega_c)
\label{eq:gamma_crit_first_model}
\end{equation} 
the largest real part of the roots becomes positive. Later, we will show that this leads to the violation of the stability criterion $\Omega^2_\text{obs}(\infty)>0$.

In those cases, when the parameters $\gamma$ and $\Omega_c$ are fixed, one needs to solve this cubic equation only once. Therefore, the best general strategy is to consider the roots as elementary functions of the system's parameters and use Vieta's formulas
\begin{eqnarray}
z_1+z_2+z_3=-\Omega_c, \label{eq:vieta_1} \\
z_1 z_2+z_2 z_3+z_3 z_1=\Omega^2,  \label{eq:vieta_2} \\
z_1 z_2 z_3=-\left(\Omega^2\Omega_c-2\gamma\Omega_c^2\right) \label{eq:vieta_3}
\end{eqnarray} 
whenever is possible. From Eq.(\ref{eq:u_after_laplace}) we can express $\tilde{u}(z)$ as
\begin{equation}
\tilde{u}(z)=\frac{(\Omega_c+z)(u'(0)+z u(0))}{(z-z_1)(z-z_2)(z-z_3)}.
\end{equation}
After decomposing the right-hand side in terms of partial fractions, one can perform the inverse Laplace transformation to obtain
\begin{equation}
u(t)=-\frac{(\Omega_c+z_1)(u'(0)+z u(0))}{(z_1-z_2)(z_3-z_1)}e^{z_1 t} +\mathrm{Cycl}.
\label{eq:u_tranformed_back}
\end{equation}
Here we used the shorthand notation ``$\mathrm{Cycl}.$'' for those terms which can be obtained from the shown one by the cyclic permutations of the roots $z_1 \to z_2 \to z_3 \to z_1$ as
\begin{eqnarray}
f(z_1,z_2,z_3)+\mathrm{Cycl}.&\equiv& f(z_1,z_2,z_3)+f(z_2,z_3,z_1)+ \nonumber \\
&& +f(z_3,z_1,z_2). \label{eq:id_for_cycl}
\end{eqnarray}
Making use of the boundary conditions in \eqref{eq:bound_cond_for_u1} together with \eqref{eq:u_tranformed_back} one can calculate and fix the constants $u(0)$ and $u'(0)$. Inserting these values back into \eqref{eq:u_tranformed_back}, the function $u_1$ can be expressed as
\begin{equation}
u_1(s)=\frac{(e^{z_1 t+z_2 s}-e^{z_1 s+z_2 t})(\Omega_c+z_1)(\Omega_c+z_2)+ \textrm{Cycl.}}{ e^{z_1 t}(\Omega_c+z_1)(z_2-z_3) +\textrm{Cycl.}}.
\label{eq:result_for_u1}
\end{equation}
In a similar way one can determine $u_2(s)$, which is:
\begin{equation}
u_2(s)=\frac{ e^{z_1 s}(\Omega_c+z_1)(z_2-z_3) +\textrm{Cycl.}}{ e^{z_1 t}(\Omega_c+z_1)(z_2-z_3) +\textrm{Cycl.}}.
\label{eq:result_for_u2}
\end{equation}
The same method is applied to the inhomogeneous equation \eqref{eq:diff_eqs_for_G}. The solution with the boundary condition \eqref{eq:bound_cond_for_G1} reads 
\begin{equation}
G_1(s,\tau)=\Theta(\tau)\Theta(s-\tau)\tilde{G}_1(s,\tau),
\label{eq:g1_structure}
\end{equation}
where $\Theta(s)$ is the Heaviside function and the smooth part $\tilde{G}_1(s,\tau)\equiv \tilde{G}_1(s-\tau)$ is given by
\begin{equation}
\tilde{G}_1(s,\tau)=-\frac{\left(e^{z_1(s-\tau)}(\Omega_c+z_1)(z_2-z_3)+\mathrm{Cycl}.\right)}{(z_1 - z_2) (z_2 - z_3) (z_3 - z_1)}.
\label{eq:g1_structure_spec}
\end{equation}
Despite the simplicity of $G_1(s,\tau)$ the expression for $G_2(s,\tau)$ is much more complicated:
\begin{equation}
G_2(s,\tau)=\Theta(\tau)\Theta(s-\tau)\tilde{G}_1(s,\tau)+\Theta(\tau)\Theta(t-\tau)\tilde{G}_2(s,\tau),
\label{eq:g2_structure}
\end{equation}
\begin{equation}
\tilde{G}_2(s,\tau)=\frac{\tilde{G}_{2\text{num}}(s,\tau)}{\tilde{G}_{2\text{den}}(t)}.
\label{eq:def_G2_tilda}
\end{equation}
The denominator in (\ref{eq:def_G2_tilda}) is given by
\begin{eqnarray}
&&\tilde{G}_{2\text{den}}(t)=(z_1 - z_2) (z_2 - z_3) (z_3 - z_1) 	\nonumber\\
&&\times\left(e^{(z_1+z_2)t}(\Omega_c+z_1)(\Omega_c+z_2)(z_1-z_2)+\mathrm{Cycl.}\right),\qquad
\label{eq:denom_of_G2_tilda}
\end{eqnarray}
while the numerator can be expressed as 
\begin{widetext}
\begin{eqnarray}
\tilde{G}_{2\text{num}}(s,\tau)&=& \biggl[ e^{(z_1+z_2)t}(\Omega_c+z_1)(\Omega_c+z_2)(z_1-z_2)\biggl( e^{z_2(s-\tau)}(\Omega_c+z_2)(z_3-z_1)+e^{z_1(s-\tau)}(\Omega_c+z_1)(z_2-z_3)\biggr.\nonumber \\
&&\biggl.\biggl.\qquad- e^{z_3s-z_1\tau}(\Omega_c+z_3)(z_2-z_3)- e^{z_3s-z_2\tau}(\Omega_c+z_3)(z_3-z_1)\biggr)+\textrm{Cycl}. \biggr].
\end{eqnarray}
\end{widetext}
It is worth noting that the forms of $G_1(s,\tau)$, $G_2(s,\tau)$ as given in Eqs. \eqref{eq:g1_structure} and \eqref{eq:g2_structure} are also valid for general spectral densities as well. This is proved by using the translation property of the inverse Laplace transform \cite{Davies}. We derive this result in Appendix \ref{app:structures_of_G1_G2}.

In Refs. \cite{Ford, Fleming} the solution to \eqref{eq:diff_eqs_for_q(t)} was written in the following form (we use our notation):
\begin{equation}
    q(s)=M q(0)\dot{G}(s)+p(0)G(s) +\int_0^s d\lambda \, G(s-\lambda)f(\lambda), 
\end{equation}
for $s \in (0,t)$, where this $G$ fulfills the homogeneous, generalized Langevin equation with boundary conditions $G(0)=0$ and $\dot{G}(0)=1/M$. Despite the different formalisms our solution is the same in the time window  $(0,t)$ due to the equations
\begin{eqnarray}
\tilde{G}_1(s,\tau)\equiv \tilde{G}_1(s-\tau)=G(s-\tau), \\
\left( u_1(s)-\frac{\dot{u}_1(0)}{\dot{u}_2(0)} u_2(s)\right)=M\dot{G}(s), \\
\frac{u_2(s)}{\dot{u}_2(0)}=M G(s).
\end{eqnarray}
In other words, the smooth part of the first Green's function $\tilde{G}_1$ in our approach is the same as theirs. However, the Heaviside functions in Eqs. \eqref{eq:g1_structure} and \eqref{eq:g2_structure} play an important role: they determine the bounds of single and triple integrals for the coefficients of the master equation (see Eqs. \eqref{eq:def_of_c(t)} and \eqref{eq:def_of_d(t)}). We devote Appendix \ref{app:Careful_evaluation_triple_integrals} to this question and show that we have only causal contributions to $C(t)$ and $D(t)$. 

Now, we have the explicit solutions of $u_1$, $u_2$, $G_1$, $G_2$ and thus time-dependent coefficients can be determined, which will be subject of our next subsection.  

\subsection{Calculation of $A(t)$, $B(t)$, $C(t)$, $D(t)$}
\label{subsec:technical_details}

By using Eqs. \eqref{eq:def_of_a(t)}, \eqref{eq:def_of_b(t)}, \eqref{eq:result_for_u1}, and \eqref{eq:result_for_u1} with \eqref{eq:result_for_eta} a direct calculation promptly leads to the exact coefficients $A(t)$ and $B(t)$ for Ohmic spectral density with a Lorentz-Drude type function and
a high-frequency cutoff \eqref{eq:Ohmic_spectral_density}:
\begin{equation}
A(t)= \frac{2M\gamma\Omega_c^2\Bigl(e^{(z_1+z_2)t}(z_1-z_2)z_3+\textrm{Cycl.}\Bigr)}{e^{(z_1+z_2)t}(z_1-z_2)(\Omega_c+z_1)(\Omega_c+z_2)+\textrm{Cycl.}},
\label{eq:result_for_a(t)}
\end{equation}
and
\begin{equation}
B(t)= \frac{2\gamma\Omega_c^2\Bigl(e^{(z_1+z_2)t}(z_1-z_2)+\textrm{Cycl.}\Bigr)}{e^{(z_1+z_2)t}(z_1-z_2)(\Omega_c+z_1)(\Omega_c+z_2)+\textrm{Cycl.}},
\label{eq:result_for_b(t)}
\end{equation}
where we have used \eqref{eq:vieta_1}.

Let us define the linear operators acting on a function $r(t)$ with one argument as
\begin{eqnarray}
\hat{C}_1 r(t)&=&\int_0^\infty d\lambda\, G_1(t,\lambda)\,r(t-\lambda), \\
\hat{C}_3 r(t)&=&\tilde{G}_{2\text{den}}(t) \int_0^t ds \int_0^\infty d\tau \int_0^\infty d\lambda \, \eta(t-s) \nonumber \\ &&\times G_1(t,\lambda)  G_2(s,\tau)\, r(\tau-\lambda), \\
\hat{D}_1 r(t)&=&\int_0^\infty d\lambda\, G'_1(t,\lambda)\,r(t-\lambda), \\
\hat{D}_3 r(t)&=&\tilde{G}_{2\text{den}}(t) \int_0^t ds \int_0^\infty d\tau \int_0^\infty d\lambda \, \eta(t-s) \nonumber \\ &&\times G'_1(t,\lambda)  G_2(s,\tau)\, r(\tau-\lambda). 
\end{eqnarray}
Direct comparison with Eqs. \eqref{eq:def_of_c(t)} and \eqref{eq:def_of_d(t)} shows that $C(t)$ and $D(t)$ can be expressed as
\begin{eqnarray}
C(t)&=&\frac{\hbar}{M}C_1(t)-\frac{2\hbar}{M^2}C_3(t), \\
C_1(t)&=&\hat{C}_1 \nu(t), \label{eq:C_1(t)_using_C_1_op}\\
C_3(t)&=&\frac{\hat{C}_3 \nu(t)}{\tilde{G}_{2\text{den}}(t)},
\label{eq:C3_fraction}\\
D(t)&=&\hbar D_1(t)-\frac{2\hbar}{M}D_3(t), \\
D_1(t)&=&\hat{D}_1 \nu(t), \\
D_3(t)&=&\frac{\hat{D}_3 \nu(t)}{\tilde{G}_{2\text{den}}(t)}.
\label{eq:D3_fraction}
\end{eqnarray}
The denominators in Eqs. \eqref{eq:C3_fraction} and \eqref{eq:D3_fraction} can be found in \eqref{eq:denom_of_G2_tilda}.
Acting with the above-defined operators to a single exponential function $\exp(-\alpha t)$, where $\alpha$ is an arbitrary constant, direct calculation leads to
\begin{eqnarray}
\hat{C}_1 e^{-\alpha t}=C_{1,0}(\alpha)+\sum_{k=1}^3 C_{1,k}(\alpha)e^{-\alpha  t+z_k t}, \label{eq:C_op_act_to_exponential}\\
\hat{D}_1 e^{-\alpha t}=D_{1,0}(\alpha)+\sum_{k=1}^3 D_{1,k}(\alpha)e^{-\alpha  t+z_k t}, \label{eq:D_op_act_to_exponential}
\end{eqnarray}
where the coefficients $C_{1,0},C_{1,k},D_{1,0},D_{1,k}$ do not depend on $t$. Explicit forms for these coefficients can be found in Appendix \ref{app:exp_ceffs}.

By writing $\cos(\omega s)=(e^{i\omega s}+e^{-i\omega s})/2$ in \eqref{eq:def_nu_at_T_eq_zero} the 
$s$ dependence occurs in the exponents, thus we can use directly the useful formulas in \eqref{eq:C_1(t)_using_C_1_op} and \eqref{eq:C_op_act_to_exponential} with $\alpha=\mp i\omega$:
\begin{eqnarray}
&&C_1 (t)=\int_0^\infty d\omega\, \frac{I(\omega)}{2}\Bigl(\hat{C}_1 e^{i\omega t} +\hat{C}_1 e^{-i\omega t}\Bigr)\nonumber\\
&&\qquad =\int_0^\infty d\omega\,\frac{I(\omega)}{2} \Bigl( C_{1,0}(-i\omega)+C_{1,0}(i\omega)\Bigr) \nonumber \\
&&+ \sum_{k=1}^3 e^{z_k t}\int_0^\infty d\omega\,\frac{I(\omega)}{2} \Bigl( C_{1,k}(-i\omega)e^{i\omega t}+C_{1,k}(i\omega)e^{-i\omega t}\Bigr).\nonumber \\ \label{eq:how_to_calculate_C_1(t)} 
\end{eqnarray}
In a similar way one gets
\begin{eqnarray}
&&\qquad D_1 (t) =\int_0^\infty d\omega\,\frac{I(\omega)}{2} \Bigl(D_{1,0}(-i\omega)+D_{1,0}(i\omega)\Bigl) \nonumber \\
&&+ \sum_{k=1}^3 e^{z_k t}\int_0^\infty d\omega\,\frac{I(\omega)}{2} \Bigl( D_{1,k}(-i\omega)e^{i\omega t}+D_{1,k}(i\omega)e^{-i\omega t}\Bigr). \nonumber \\\label{eq:how_to_calculate_D_1(t)}
\end{eqnarray}
The integration over $\omega$ is either elementary or can be expressed in terms of $I_1(r,t)$ or $I_2(r,t)$, which are introduced in Appendix~\ref{app:useful_integrals}, where the parameter $r$ takes values of $\Omega_c$, $z_1$, $z_2$ and $z_3$.

The same idea works for those parts of the coefficients $C(t)$ and $D(t)$ which are related to the triple integrals in \eqref{eq:def_of_c(t)} and \eqref{eq:def_of_d(t)}.
The actions on an exponential function of operators $\hat{C}_3$ and $\hat{D}_3$ require much more work. They can be summarized as follows. Now, instead of \eqref{eq:C_op_act_to_exponential} and \eqref{eq:D_op_act_to_exponential} we have 
\begin{eqnarray}
&&\hat{C}_3 e^{-\alpha t}=\sum_{i=1}^6 C_{3,i}(\alpha)e^{\epsilon_i t}+\sum_{i=7}^{13} C_{3,i}(\alpha)e^{\epsilon_i t-\alpha t},\quad \label{eq:C_3_op_act_to_exponential}\\
&&\hat{D}_3 e^{-\alpha t}=\sum_{i=1}^6 D_{3,i}(\alpha)e^{\epsilon_i t}+\sum_{i=7}^{13} D_{3,i}(\alpha)e^{\epsilon_i t-\alpha t},\quad 
\label{eq:D_3_op_act_to_exponential}
\end{eqnarray} 
where the exponents $\epsilon_i$ are enumerated in the following table:
\[
\begin{tabular}{|r|c||r|c|}
\hline
i & $\epsilon_i$ & i & $\epsilon_i$ \\
\hline
1 & $2z_1+z_2+z_3$ & 7 & $2z_1+z_2$ \\
\hline
2 & $z_1+2z_2+z_3$ & 8 & $2z_2+z_3$ \\
\hline
3 & $z_1+z_2+2z_3$ & 9 & $2z_3+z_1$\\
\hline
4 & $z_1+z_2$ & 10 & $2z_1+z_3$ \\
\hline
5 & $z_2+z_3$ & 11 & $2z_2+z_1$ \\
\hline
6 & $z_3+z_1$ & 12 & $2z_3+z_2$  \\
\hline
& & 13 & $z_1+z_2+z_3$ \\
\hline
\end{tabular}.
\] 
Formulas for the coefficients appearing in the right-hand sides of Eqs. (\ref{eq:C_3_op_act_to_exponential}), (\ref{eq:D_3_op_act_to_exponential}) are shown in Appendix \ref{app:exp_ceffs}. In a similar way as we got (\ref{eq:how_to_calculate_C_1(t)}) and (\ref{eq:how_to_calculate_D_1(t)}) $C_3(t)$ and $D_3(t)$, can be calculated via
\begin{widetext}
\begin{eqnarray}
&&C_3 (t)=\frac{1}{\tilde{G}_{2\text{den}}(t)}\int_0^\infty d\omega\, \frac{I(\omega)}{2}\left(\hat{C}_3 e^{i\omega t} +\hat{C}_3 e^{-i\omega t}\right)\nonumber\\
&&=\frac{1}{\tilde{G}_{2\text{den}}(t)}\left[\sum_{i=1}^6 e^{\epsilon_it}\int_0^\infty d\omega\,\frac{I(\omega)}{2} \Bigl(C_{3,i}(-i\omega)+C_{3,i}(i\omega)\Bigr)+ \sum_{i=7}^{13} e^{\epsilon_i t}\int_0^\infty d\omega\,\frac{I(\omega)}{2} \Bigl( C_{3,i}(-i\omega)e^{i\omega t}+C_{3,i}(i\omega)e^{-i\omega t}\Bigr)\right],  \nonumber \\ \label{eq:how_to_calculate_C_3(t)} \\
&&D_3 (t)=\frac{1}{\tilde{G}_{2\text{den}}(t)}\int_0^\infty d\omega\, \frac{I(\omega)}{2}\left(\hat{D}_3 e^{i\omega t} +\hat{D}_3 e^{-i\omega t}\right)\nonumber\\
&&=\frac{1}{\tilde{G}_{2\text{den}}(t)}\left[\sum_{i=1}^6 e^{\epsilon_it}\int_0^\infty d\omega\,\frac{I(\omega)}{2} \Bigl(D_{3,i}(-i\omega)+D_{3,i}(i\omega)\Bigr)+ \sum_{i=7}^{13} e^{\epsilon_i t}\int_0^\infty d\omega\,\frac{I(\omega)}{2} \Bigl( D_{3,i}(-i\omega)e^{i\omega t}+D_{3,i}(i\omega)e^{-i\omega t}\Bigr)\right].  \nonumber \\ \label{eq:how_to_calculate_D_3(t)}
\end{eqnarray}
\end{widetext}
Once again we have the opportunity to express all the necessary $\omega$ integrals in terms of $I_1(r,t)$ or $I_2(r,t)$ with $r=\Omega_c,z_1,z_2,z_3$.

Let us introduce the following vectors ($T$ denotes the transpose):
\begin{eqnarray}
\mathbf{v}_1(t)&=&(e^{z_1 t},e^{z_2 t},e^{z_3 t})^T, \\
\mathbf{v}_2(t)&=&(e^{\varepsilon_1 t},\ldots,e^{\varepsilon_6 t} )^T, \\
\mathbf{v}_3(t)&=&(e^{\varepsilon_7 t},\ldots,e^{\varepsilon_{13} t} )^T,\\
\mathbf{v}_4&=&(\ln(\Omega_c^2),\ln{(z_1^2)},\ln{(z_2^2)},\ln{(z_3^2)})^T, \\
\mathbf{v}_5(t)&=&(I_1(\Omega_c,t),I_1(z_1,t),I_1(z_2,t),I_1(z_3,t), \nonumber \\
&& \quad I_2(\Omega_c,t),I_2(z_1,t),I_2(z_2,t),I_2(z_3,t))^T.
\end{eqnarray}
The full time dependence of the coefficients can be summarized as
\begin{eqnarray}
C_1(t)&=&C_1(\infty)+\mathbf{v}_1^T(t)\cdot \mathbf{M}_{C1}\cdot \mathbf{v}_5(t), \\
D_1(t)&=&D_1(\infty)+\mathbf{v}_1^T(t)\cdot \mathbf{M}_{D1}\cdot \mathbf{v}_5(t), \\
C_3(t)&=&\frac{\mathbf{v}_2^T(t) \cdot \mathbf{M}_{C31} \cdot \mathbf{v}_4+ \mathbf{v}_3^T (t) \cdot \mathbf{M}_{C32} \cdot \mathbf{v}_5(t)}{\tilde{G}_{\text{2den}}(t)}, \quad \\
D_3(t)&=&\frac{\mathbf{v}_2^T(t) \cdot \mathbf{M}_{D31} \cdot \mathbf{v}_4+ \mathbf{v}_3^T (t) \cdot \mathbf{M}_{D32} \cdot \mathbf{v}_5(t)}{\tilde{G}_{\text{2den}}(t)},
\end{eqnarray}
where scalars $C_1(\infty)$ and $D_1(\infty)$ and the matrices $\mathbf{M}_{C1}$, $\mathbf{M}_{D1}$, $\mathbf{M}_{C31}$, $\mathbf{M}_{C32}$, $\mathbf{M}_{D31}$, $\mathbf{M}_{D32}$ are 
independent of time.

\subsection{Short-time expansions of the coefficients}
\label{subsec:short_time_expansions}

In this section, we are interested in the
behavior of the time-dependent coefficients around 
$t=0$, when the interaction between the central harmonic oscillator and the bath is switched on. A lengthy but still straightforward calculation leads to the expressions
\begin{equation}
A_s(t) \simeq -2 M\gamma\Omega_c (\Omega_c t)+ M \gamma \Omega_c (\Omega_c t)^2+ O(t^3),
\label{eq:A(t)_short_time},
\end{equation}
and
\begin{equation}
B_s(t) \simeq  \gamma  (\Omega_c t)^2+ O(t^3).
\label{eq:B(t)_short_time}
\end{equation}
In $C(t)$ and $D(t)$ the leading order corrections, which come only from the single integrals in  (\ref{eq:def_of_c(t)}) and (\ref{eq:def_of_d(t)})) up to $O(t^3 \ln{t})$ are:
\begin{equation}
C_s(t) \simeq  \frac{\hbar}{2\pi}\gamma (\Omega_c t)^2\Bigl[1-2\gamma_{EM}-2\ln(\Omega_c t)\Bigr] +O(t^3\ln(t)),
\label{eq:C(t)_short_time}
\end{equation}
\begin{equation}
D_s(t) \simeq  \frac{2\hbar}{\pi}M\gamma \Omega_c (\Omega_c t)\Bigl[1-\gamma_{EM}-\ln(\Omega_c t)\Bigr] +O(t^3\ln(t)),
\label{eq:D(t)_short_time}
\end{equation}
where $\gamma_{EM}$ is the Euler-Mascheroni constant, which is approximately 0.577. Surprisingly, the same short time expressions were found in Appendix C of \cite{HCsCsB}. Thus, the short time behavior of the coefficients at $T=0$ is not influenced by the weak coupling limit approximation. 

\subsection{Asymptotic values}
\label{subsec:asymptots}

In the parameter region $0< \gamma< \gamma_\text{cr}= \Omega^2/(2\Omega_c)$ one can have either one real and two complex or three real negative roots
of Eq.~\eqref{eq:cubic_equation} (both possibilities occur actually). The root with the smallest real part is always real and is close to $(-\Omega_c)$ (see Fig.~\ref{fig:roots_of_orig_model}). Let us denote this root by $z_1$. If the parameters are chosen such that the coefficients converge, i.e., $0< \gamma< \gamma_\text{cr}$, one can determine the asymptotic values:
\begin{equation}
A(\infty)= - 2M\gamma \Omega_c^2\frac{(\Omega_c+z_2+z_3)}{(\Omega_c+z_2)(\Omega_c+z_3)},
\label{eq:A(t)_stac}
\end{equation}
\begin{equation}
B(\infty)=  \frac{2\gamma \Omega_c^2}{(\Omega_c+z_2)(\Omega_c+z_3)},
\label{eq:B(t)_stac}
\end{equation}
\begin{widetext}
\begin{eqnarray}
C(\infty)&=&\frac{\hbar\gamma\Omega_c^2}{\pi} \left(\frac{z_1 \ln{(z_1^2/\Omega_c^2)}}{(\Omega_c-z_1)(z_1-z_2)(z_1-z_3)}+\textrm{Cycl.}\right)+\frac{2\hbar\gamma^2 \Omega_c^4}{\pi}\Biggl( \frac{(z_2-z_3)\left[z_1^2(z_2+z_3)+\Omega_c(z_1^2+z_2z_3)\right]\ln{(z_1^2/\Omega_c^2)}}{(\Omega_c^2-z_1^2)(z_1+z_2)(z_1+z_3)}\Biggr.\nonumber\\
&& \Biggl. +\frac{z_2(z_3-z_1)\ln{(z_2^2/\Omega_c^2)}}{(\Omega_c-z_2)(z_1+z_2)}+ \frac{z_3(z_1-z_2)\ln{(z_3^2/\Omega_c^2)}}{(\Omega_c-z_3)(z_1+z_3)}\Biggr)\Bigg/ \Biggl( (z_1-z_2)(z_2-z_3)(z_3-z_1)(\Omega_c+z_2)(\Omega_c+z_3) \Biggr),
\label{eq:C(t)_stac}
\end{eqnarray}
\begin{eqnarray}
D(\infty)&=&\frac{\hbar M\gamma\Omega_c^2}{\pi} \left(\frac{z_1^2 \ln{(z_1^2/\Omega_c^2)}}{(\Omega_c-z_1)(z_1-z_2)(z_1-z_3)}+\textrm{Cycl.}\right)+\frac{2\hbar M \gamma^2 \Omega_c^4}{\pi}\Biggl( \frac{z_1^2(z_2-z_3)\left[z_1^2+z_2 z_3+\Omega_c(z_2+z_3)\right]\ln{(z_1^2/\Omega_c^2)}}{(\Omega_c^2-z_1^2)(z_1+z_2)(z_1+z_3)}\Biggr.\nonumber\\
&& \Biggl. +\frac{z_2^2(z_3-z_1)\ln{(z_2^2/\Omega_c^2)}}{(\Omega_c-z_2)(z_1+z_2)}+ \frac{z_3^2(z_1-z_2)\ln{(z_3^2/\Omega_c^2)}}{(\Omega_c-z_3)(z_1+z_3)}\Biggr)\Bigg/ \Biggl( (z_1-z_2)(z_2-z_3)(z_3-z_1)(\Omega_c+z_2)(\Omega_c+z_3) \Biggr).
\label{eq:D(t)_stac}
\end{eqnarray}
\end{widetext}
\begin{figure}
	
	\includegraphics[angle=270,width=0.45 \hsize]{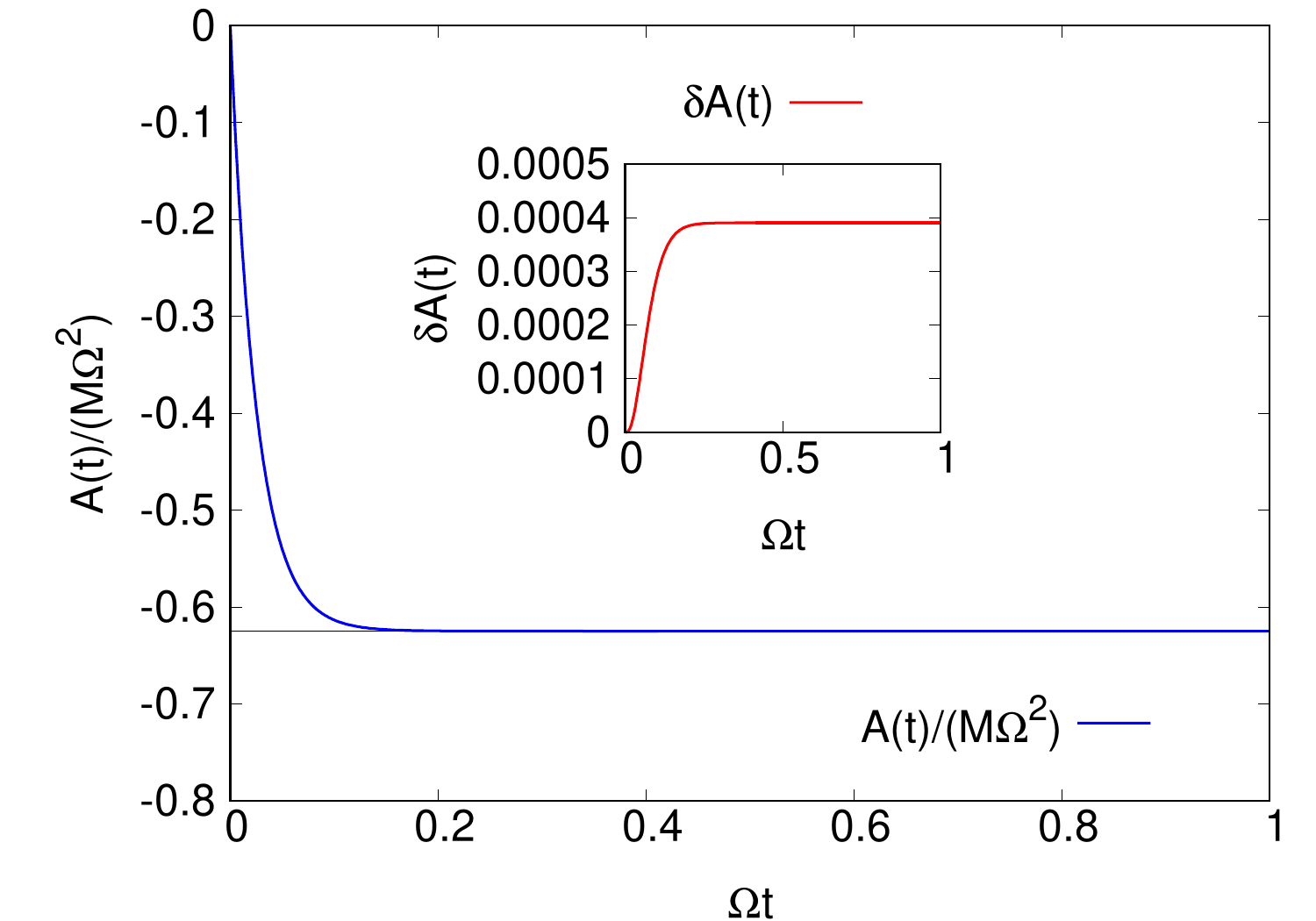} \hspace{1.5cm}
	\includegraphics[angle=270, width=0.45 \hsize]{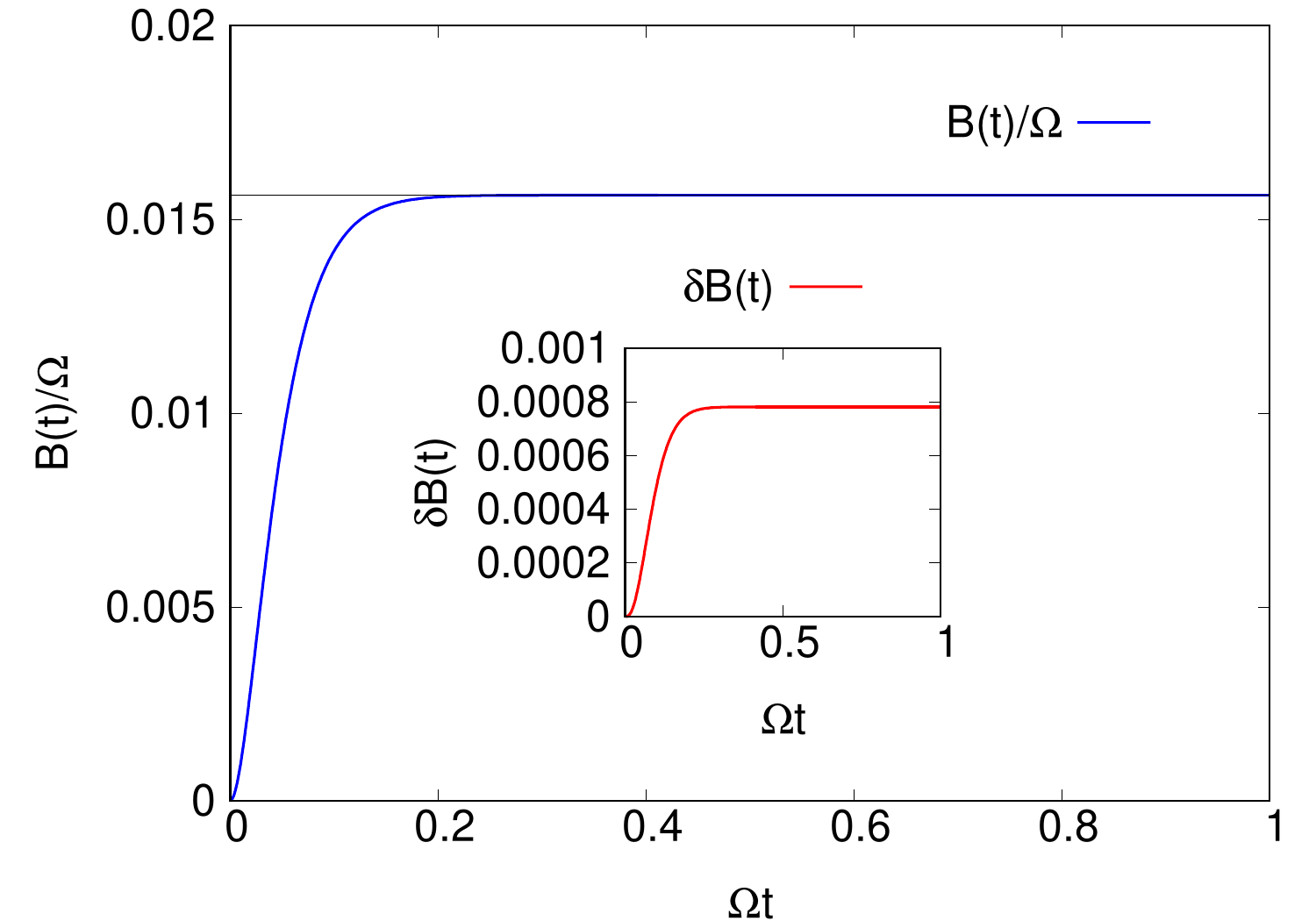}
	
	\caption{Behaviors of $A(t)$ (left) and $B(t)$ (right). Thin lines show the asymptotic values \eqref{eq:A(t)_stac} and \eqref{eq:B(t)_stac}, respectively. We set $\Omega_c=40\Omega$ and $\gamma=\Omega/128$.  In the insets, we plot the relative differences of  $\delta A(t)=2(A-A_w)/(A+A_w)$ (left) and $\delta B(t)=2(B-B_w)/(B+B_w)$ (right). On both axes, we use dimensionless units. \label{fig:A(t)_B(t)_for_normal_situation}}
\end{figure}

A direct calculation yields $\Omega^2_\text{obs}=\Omega^2+A(\infty)/M<0$, if $\gamma > \gamma_\text{crit}$. The corresponding weak coefficients are
\begin{equation}
A_w(\infty)=-2M\gamma\Omega_c^2\frac{\Omega_c}{\Omega_c^2+\Omega^2},
\end{equation}
\begin{equation}
B_w(\infty)=\frac{2\gamma\Omega_c^2}{\Omega_c^2+\Omega^2},
\end{equation}
\begin{equation}
C_w(\infty)=-\frac{2\hbar\gamma\Omega_c^2}{\pi}\frac{\ln\left(\Omega_c/\Omega\right)}{\Omega_c^2+\Omega^2},
\label{eq:C_w_asymptotic}
\end{equation}
\begin{equation}
D_w(\infty)=\frac{\hbar M\gamma\Omega_c^2 \Omega}{\Omega_c^2+\Omega^2}.
\label{eq:D_w_asymptotic}
\end{equation}
We note that $A_w(\infty)$, $B_w(\infty)$ can be obtained from \eqref{eq:A(t)_stac} and  \eqref{eq:B(t)_stac}, and $C_w(\infty)$, $D_w(\infty)$ from the first terms of \eqref{eq:C(t)_stac} and \eqref{eq:D(t)_stac}, which originate from the single integrals in \eqref{eq:def_of_c(t)} and \eqref{eq:def_of_d(t)}, where one has to evaluate 
 carefully the roots $z_1, z_2, z_3$ and the expressions for small values of $\gamma$.  The approximate roots are $z_1 \approx -\Omega_c+(\exp{(i\xi)}+\exp{(-i\xi}))\epsilon$, $z_2 \approx i\Omega-\exp{(i\xi)}\epsilon$,  $z_3 \approx -i\Omega-\exp{(-i\xi)}\epsilon$,   where $\epsilon = \gamma \Omega_c^2/(\Omega\sqrt{\Omega^2+\Omega_c^2})$ and $\exp(i\xi)=(\Omega+i\Omega_c)/\sqrt{\Omega^2+\Omega_c^2}$. These approximate roots fulfill Vieta's formulas \eqref{eq:vieta_1}-\eqref{eq:vieta_3} up to $O(\gamma)$ precision. Careful evaluations are necessary in the arguments of the logarithms in the complex plane, because $z_2^2$ and $z_3^2$ are very close to the real axis with negative real parts.   
 
\subsection{Numerical results}
\label{subsec:numerics}

For our first numerical examples, we choose reasonable values for the cutoff frequency, $\Omega_c=40\Omega$,  and $\gamma=\Omega/128$ ensuring  $\Omega_c \gg \Omega$ and $\gamma< \gamma_\text{cr}=\Omega/80$; see Eq. \eqref{eq:gamma_crit_first_model}. The first step is to solve Eq.~\eqref{eq:cubic_equation} and use $z_1,z_2,z_3$ in the corresponding formulas.
In Fig.~\ref{fig:A(t)_B(t)_for_normal_situation} we show coefficients $A(t)$ and $B(t)$, as given by Eqs.~\eqref{eq:result_for_a(t)} and \eqref{eq:result_for_b(t)} using dimensionless units.  Surprisingly, the coefficients in the weak coupling approximation $A_w(t)$ and $B_w(t)$ are close to $A(t)$ and $B(t)$.
This can be seen in the insets of Fig.~\ref{fig:A(t)_B(t)_for_normal_situation} where we show their relative differences.

Inequalities (\ref{eq:om_obs2_pos}) restrict the allowed values of  $\gamma$.  $\Omega_\text{obs}^2(\infty)$ decreases as a function of $\gamma$ and vanishes at $\gamma_\text{cr}$. For $\gamma > \gamma_\text{cr}$ the quantity  $\Omega_\text{obs}^2(\infty)$ is negative, which restricts $\gamma$ to the region 
\begin{equation}
\gamma < \gamma_\text{cr}.
\label{eq:crit_for_the_first_model}
\end{equation}
However, $A(t)$ is always independent of temperature, thus even at finite temperature  $\gamma < \gamma_\text{cr}$ should hold in the original model. This consistency condition has some interesting aspects. When the cutoff frequency $\Omega_c$ tends to infinity then $\gamma_\text{cr} \approx 0$ and $I(\omega)$ in Eq. \eqref{eq:Ohmic_spectral_density} is approximately $2M\gamma \omega /\pi$, i.e., the purely Ohmic environment. Even though the purely Ohmic environment is known to be unphysical, here in our analysis, we have been able to capture a condition whose violation leads to the problem of introducing a counterterm in the bare frequency $\Omega$ such that 
$\Omega_\text{obs}^2(\infty)$ stays positive for all times; see Ref. \cite{Kappler}.

\begin{figure*}[t!]
\begin{center}
	\includegraphics[angle=270,width=0.45 \textwidth]{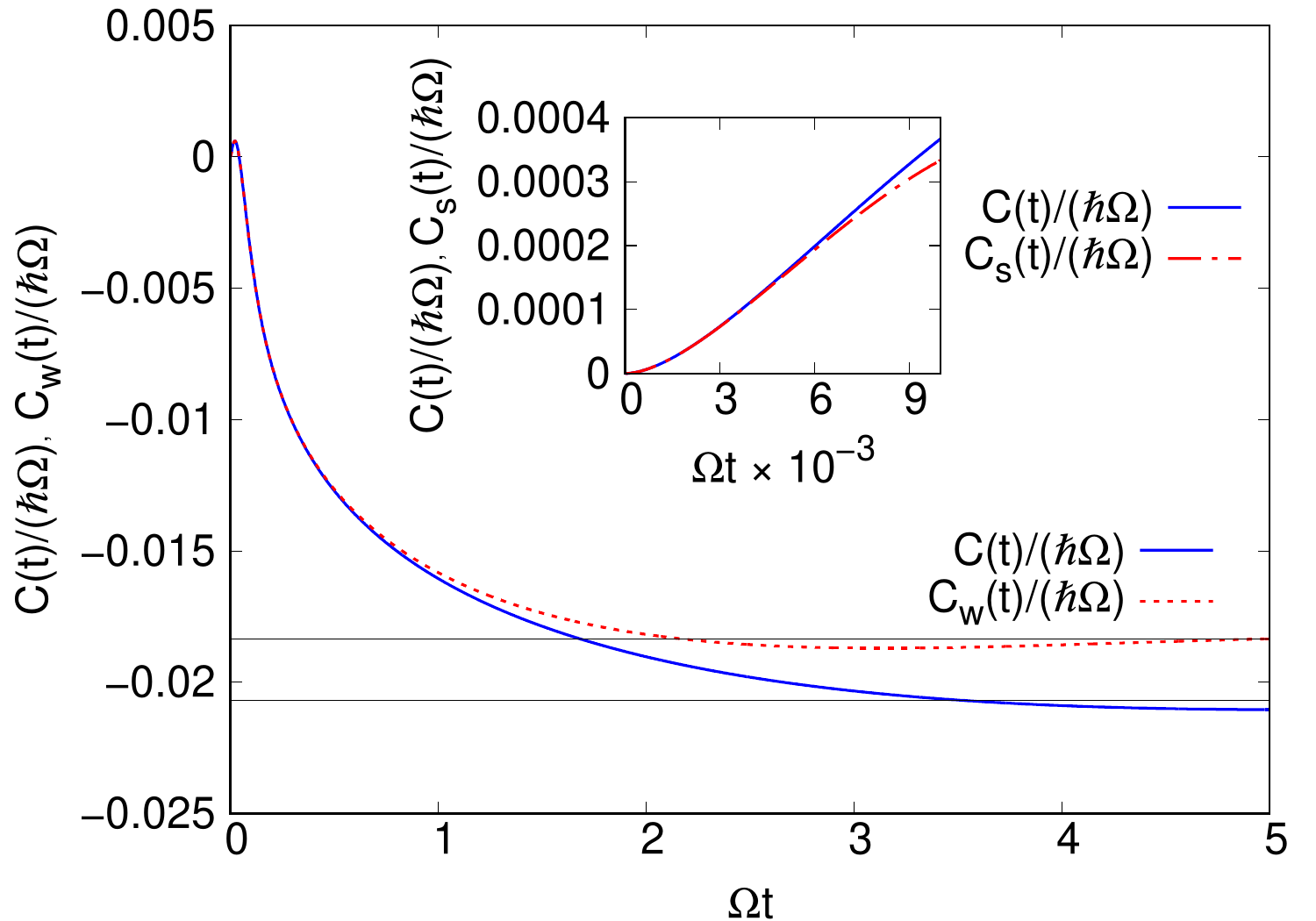} 
 \hspace{1.5cm}
	\includegraphics[angle=270,width=0.45\textwidth]{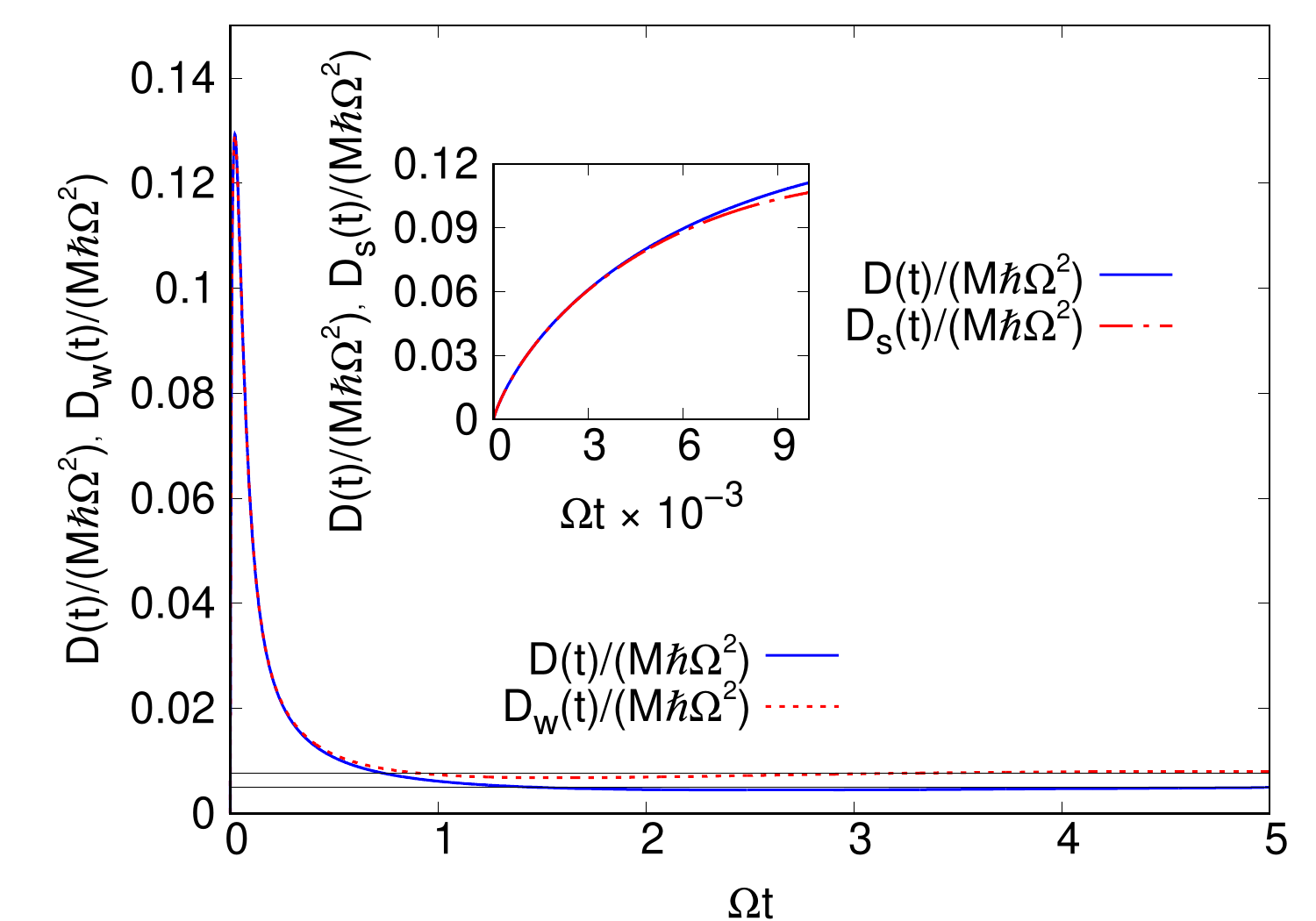}
	\caption{Behaviors of $C(t)$ (left) and $D(t)$ (right). Thin horizontal lines show the asymptotic values \eqref{eq:C(t)_stac} and \eqref{eq:D(t)_stac} and the asymptotics of the weak coupling limit  (\ref{eq:C_w_asymptotic}) and (\ref{eq:D_w_asymptotic}) , respectively. We set $\Omega_c=40\Omega$ and $\gamma=\Omega/128$. In the insets, we show the short-time behavior of $C(t)$ with the expansion $C_s(t)$ given by Eq. \eqref{eq:C(t)_short_time} (left inset) and $D(t)$ with the expansion $D_s(t)$ given by \eqref{eq:D(t)_short_time} (right inset), respectively. On both axes, we use dimensionless units. \label{fig:C(t)_D(t)_for_normal_situation}}
\end{center}
\end{figure*}

In Fig.~\ref{fig:C(t)_D(t)_for_normal_situation} we show the coefficients $C(t)$ and $D(t)$ for the same parameters used in Fig.~\ref{fig:A(t)_B(t)_for_normal_situation}. It is important to note that the so-called initial
jolt discussed by Ref. \cite{HuPazZhang92, Unruh} appears, as expected. However, this peak is developed only for a short timescale of the bare frequency $\Omega$.  Furthermore, according to the depicted curves, the exact result and the result of the weak coupling limit differs considerably over a long time. These big differences can be also seen from the different asymptotic values $C(\infty)$ and $C_w(\infty)$ (or in $D(\infty)$ and $D_w(\infty)$) and can be attributed to the triple integrals of Eqs.~\eqref{eq:def_of_c(t)}  and \eqref{eq:def_of_d(t)}. These integrals are not taken into account in the weak coupling limit.  The situation for very short times is different: $C(t)$ and $C_w(t)$ (or $D(\infty)$ and $D_w(\infty)$) differs only a little.  From the insets in Fig.~\ref{fig:C(t)_D(t)_for_normal_situation} it is clear that \eqref{eq:C(t)_short_time} and  \eqref{eq:D(t)_short_time} are the leading behavior for very short times. We note that up to the orders considered the triple integrals do not give any contribution for very short times.  

For fixed $\Omega_c$ and $\gamma<\gamma_\text{cr}$ the coefficients $A(t)$, $B(t)$, $C(t)$ and $D(t)$ tend to constants; see Sec.~\ref{subsec:asymptots}. Thus, the master equation is asymptotically Markovian, which was discussed in Sec. \ref{I.and.half}. Inequality  (\ref{eq:Q_def}) (i.e., $Q \geq 1$) necessarily holds for a physical system.

\begin{figure}[hb!]
	\[
	\includegraphics[angle=270, width=0.9 \hsize]{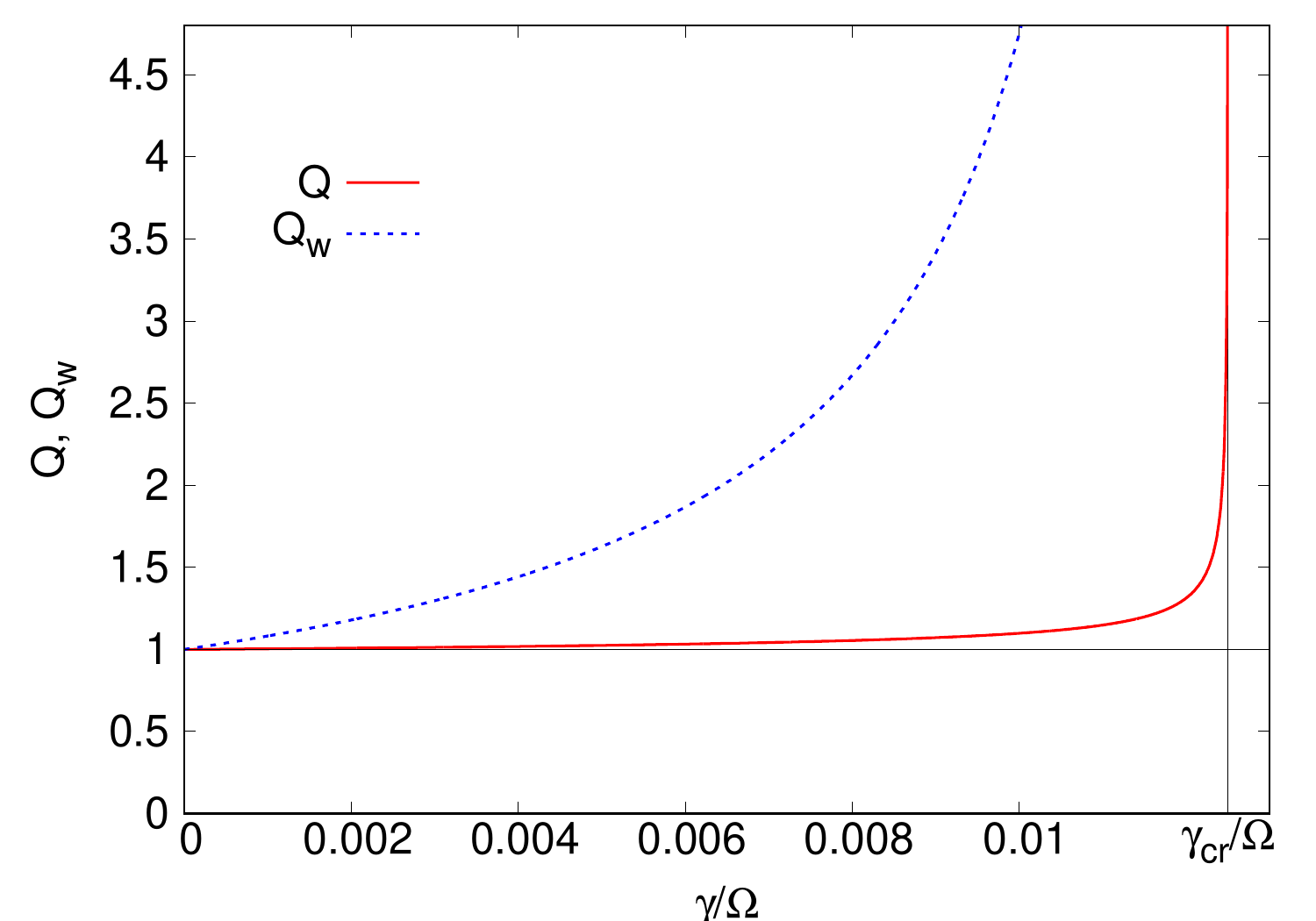} 	
	\]
	\caption{$Q$ and $Q_w$ calculated from the stationary values of coefficients of the non-Markovian master equation as a function of $\gamma$. The continuous line shows $Q$; dashed line is for $Q_w$. We set $\Omega_c=40\Omega$. On the axes, we use dimensionless quantities.\label{fig:a_per_c}}
\end{figure}

Replacements $A(\infty) \to A_w(\infty)$, etc., define the quantity $Q_w$ for the weak coupling limit. In Fig~\ref{fig:a_per_c} we show both $Q$ and $Q_w$ as a function of $\gamma$, and find that $Q$ and $Q_w$ are bigger than $1$ in the region $0 \le \gamma < \gamma_\text{cr}$. 
This figure also shows that the weak coupling limit does not necessarily mean that $Q$ are closely approximated by $Q_w$ for small values of $\gamma$. On the other hand, 2$\gamma$ is approximately equal to the damping constant $B(\infty)$
of the central oscillator as $t \to \infty$ and $\Omega_c \gg \Omega$. Thus, in general, weak damping and the weak coupling limit are not related to each other.

\section{Comparison of three models}
\label{IV}

In this section, we compare the following models.  
The first model is given by the Hamiltonian in \eqref{eq:full_hamiltonian} analyzed in the preceding sections. The second is called in the literature the Caldeira-Leggett model \cite{Grabert} and
\cite{CL}. It differs from the Hamiltonian of the first model \eqref{eq:full_hamiltonian} in the coupling:
\begin{eqnarray}
\hat{H}_\text{CL}&=&\frac{\hat{p}^2}{2M} +\frac12 M\Omega^2 \hat{q}^2 \nonumber \\
&& +\sum_n \left[\frac{\hat{p}_n^2}{2m_n}+\frac12 m_n \omega_n^2\left( \hat{q}_n + \frac{C_n}{m_n \omega_n^2}\hat{q}\right)^2\right]=\nonumber \\
&&=\hat{H}_{\text{orig}}+\frac12 \hat{q}^2 \sum_n \frac{C_n^2}{m_n \omega_n^2}.
\label{eq:second_hamiltonian}
\end{eqnarray} 
In the original Caldeira-Leggett model $(-C_n)$ stands for $C_n$; however, the coefficients of the reduced master equation are insensitive to this sign change.

The sum can be expressed by the spectral density and reads
\begin{equation}
\sum_n \frac{C_n^2}{m_n \omega_n^2}=2 \int_0^\infty d\omega\, \frac{I(\omega)}{\omega}=2 \gamma\Omega_c.
\end{equation} 
The new term effectively shifts the bare frequency of the central harmonic oscillator as $\Omega^2 \to \Omega^2+2\gamma\Omega_c$. All the properties of the second model can be derived from the corresponding results of the first model by applying this replacement everywhere.

\begin{figure}[hb!]
	\includegraphics[angle=0,width=0.9 \hsize]{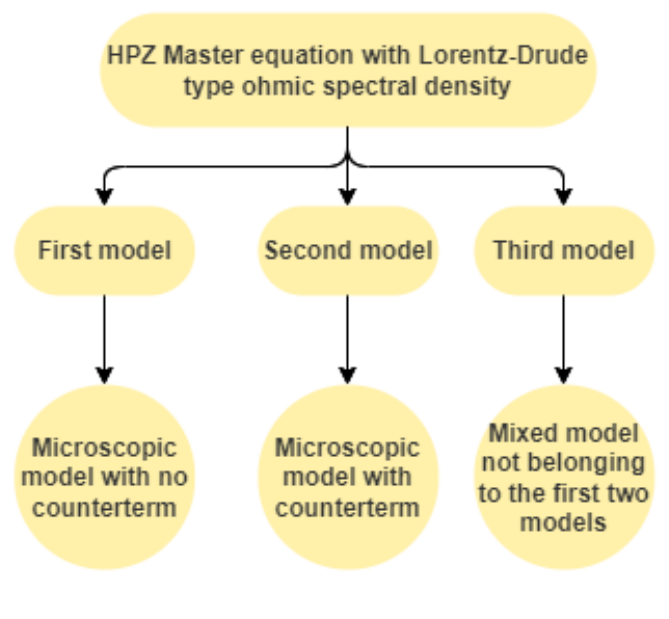}
	\caption{Schematic diagram  which shows the
differences between the three models.
		\label{fig:schematic_diagram}}
\end{figure}

A third group of models consists of those which are not of the first or second type. See the diagram in Fig.~\ref{fig:schematic_diagram}.  Some examples are in Refs. \cite{book1,HCsCsB}, where a counterterm is  used 
in the weak approximation with the following strategy:  the replacement $\Omega^2 \to \Omega^2+2\gamma\Omega_c$ is used in the kernels $\eta(s)$ and $\nu(s)$, but not in the expressions $\cos(\Omega s)$ and $\sin(\Omega s)$  in Eqs.~\eqref{eq:a(t)_in_weak_coupling}, \eqref{eq:b(t)_in_weak_coupling}, \eqref{eq:c(t)_in_weak_coupling}, and \eqref{eq:d(t)_in_weak_coupling}. This means that the resulting model is not directly connected to the original microscopic model,  because the shift was not considered everywhere where $\Omega$ occurred in the first model. 

\begin{figure}
	\includegraphics[angle=270,width=0.9 \hsize]{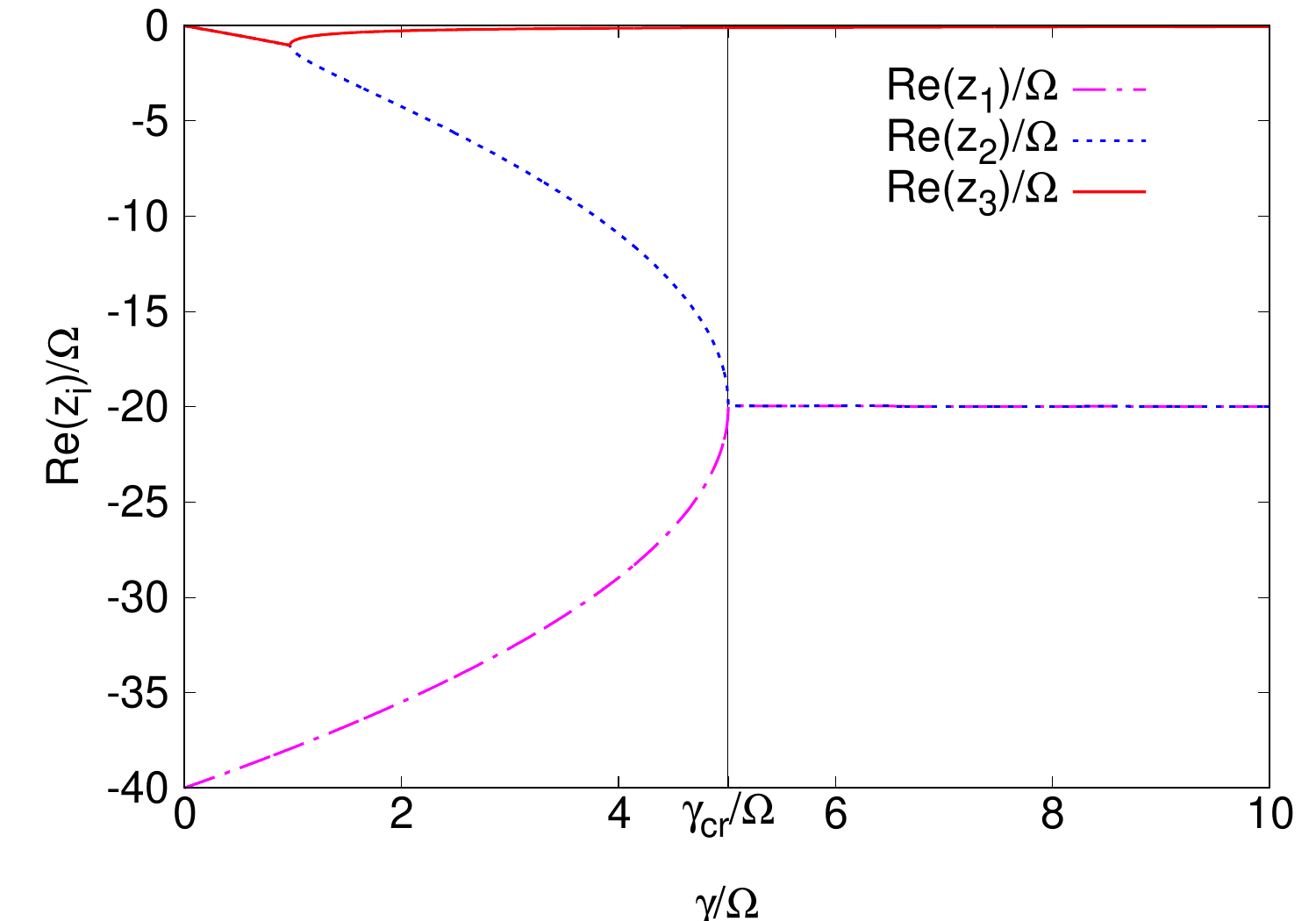}
	\caption{ Real parts of the three roots of Eq. (\ref{eq:new_cubic_equation}) as a function of the coupling parameter $\gamma$ at fixed $\Omega_c=40\Omega$. On the axes, we use dimensionless quantities. The vertical line indicates the new critical coupling $\gamma_{\text{cr}}$.\label{fig:roots_of_renorm_model}}
\end{figure}

In the case of the second model, the shift affects also the structure of the three roots. Instead of Eq.~\eqref{eq:cubic_equation} one has to solve 
\begin{equation}
z^3+\Omega_c z^2+(\Omega^2 +2\gamma\Omega_c)z+\Omega^2\Omega_c=0.
\label{eq:new_cubic_equation}
\end{equation}
For the same values of the parameters $\Omega,\Omega_c,\gamma$ the tree roots differs considerably; compare Fig.~\ref{fig:roots_of_renorm_model} with Fig.~\ref{fig:roots_of_orig_model}.

\begin{figure}
	\includegraphics[angle=270,width=0.9 \hsize]{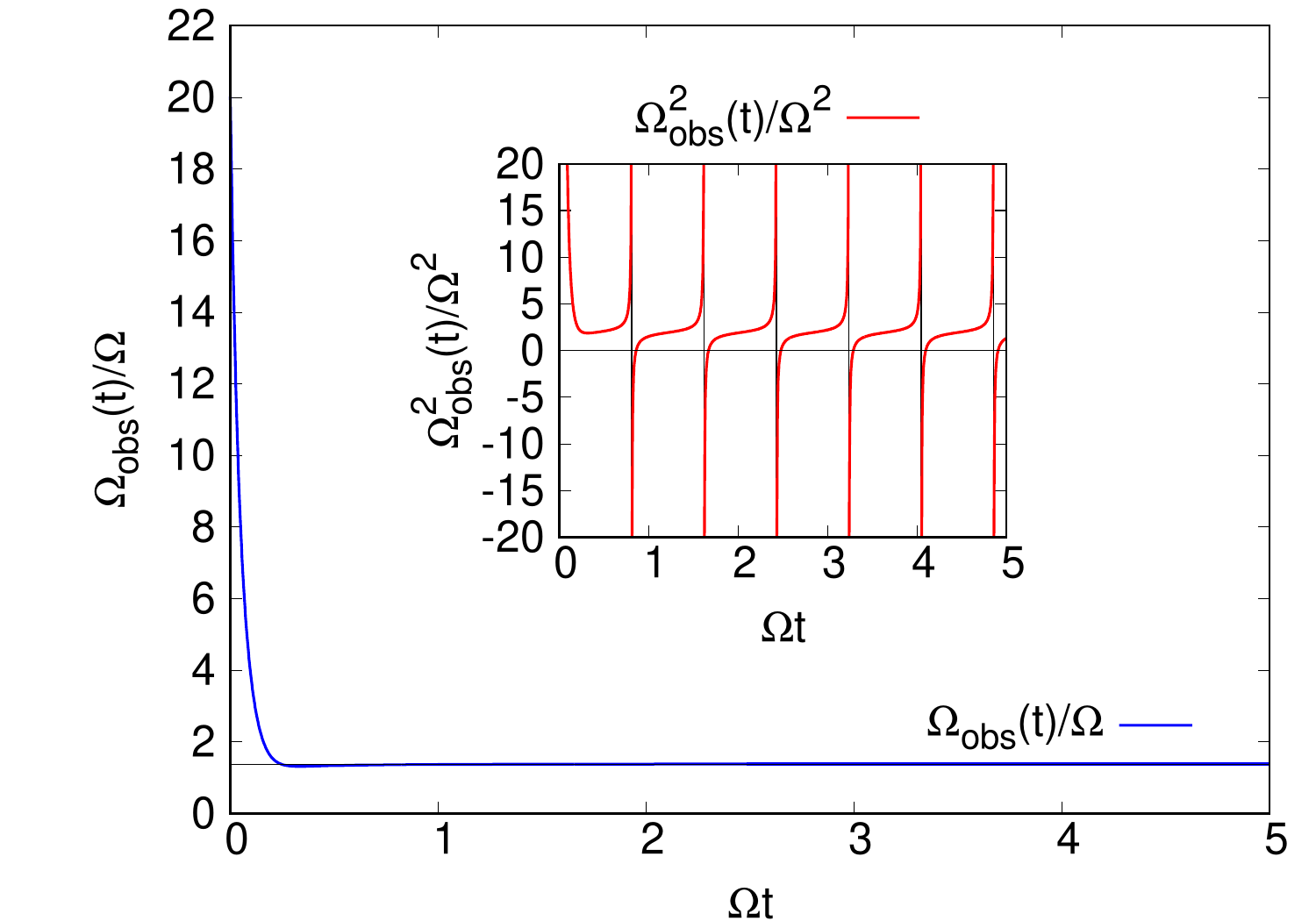}
	\caption{ The observed frequency $\Omega_\text{obs}(t)$ given by (\ref{eq:observed_freq_for_the_second_model}) as a function of $t$ for $\Omega_c=40\Omega$ and $\gamma=5\Omega < \gamma_{\text{cr}}=5.01253\,\Omega$. The inset shows the squared $\Omega_{\text{obs}}^2$ as a function of $t$, but for $\Omega_c=40\Omega$ and $\gamma=5.2\Omega >\gamma_{\text{cr}}$. On the axes we use dimensionless quantities.
		\label{fig:new_observed_omega}}
\end{figure}

Now, the real parts of the three roots of Eq. \eqref{eq:new_cubic_equation} are always negative. However, a strange phenomenon appears at a new critical coupling (indicated by $\gamma_{\text{cr}}$ in Fig.~\ref{fig:roots_of_renorm_model}), where out the three real roots two become conjugated complex by increasing $\gamma$. This is exactly the opposite phenomenon that we observed for the first model.
Due to the shift $2\gamma \Omega_c$  the quantity $\Omega_\text{obs}^2(t)$ should be defined  according to
\begin{equation}
\Omega_\text{obs}^2(t)=\Omega^2+2\Omega_c\gamma+\frac{A(t)}{M}.
\label{eq:observed_freq_for_the_second_model}
\end{equation}
Above $\gamma_{\text{cr}}$ the observed physical frequency  square of this model
diverges periodically as a function of time. This is shown in Fig.~\ref{fig:new_observed_omega}. This problem arises independently of the temperature. Thus, this model is also nonphysical for fixed $\Omega_c$ as soon as $\gamma > \gamma_{\text{cr}}$ at any temperature.

\begin{figure}
	\[
	\includegraphics[angle=270, width=0.9 \hsize]{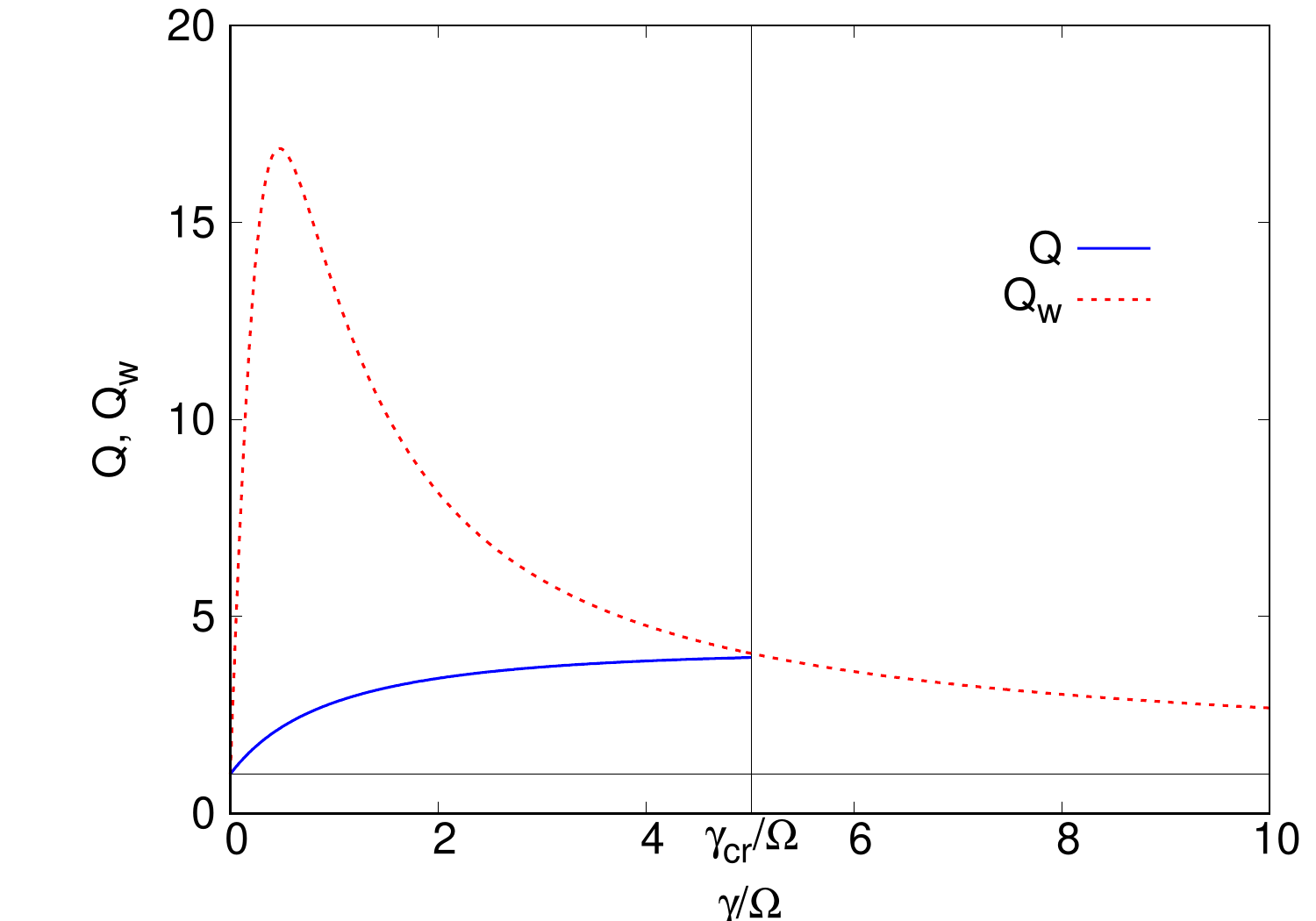} 	
	\]
	\caption{$Q$ and $Q_w$ for the second model as a function of $\gamma$. The continuous line shows Q; the dashed line is for $Q_w$. Parameter $\Omega_c=40\Omega$. On the axes we use dimensionless quantities. \label{fig:renormalt_a_per_c}}
	\label{fig:positivity_second_model}
\end{figure}

 For the first model Eq. \eqref{eq:crit_for_the_first_model} restricts $\gamma$ to a given parameter range. A useful feature of the second model is, that contrary to the first model $\gamma_\text{cr}$ is increasing with $\Omega_c$ allowing a much broader consistency range for $\gamma$. The positivity of the asymptotic state for the second model is studied in Fig.~	\ref{fig:positivity_second_model}. Parameters $Q$ and $Q_w$ are calculated from \eqref{eq:Q_def} with the shift in $\Omega$ and they differ considerably. {\it Formally}, $Q_w$ can be continued beyond $\gamma_\text{cr}$, as no singularity occurs at this point in $Q_w$, but $Q$ stops there. We stress again that beyond the critical coupling $\gamma_\text{cr}$ the model looses its validity due to the problem shown in inset of Fig.~\ref{fig:new_observed_omega}. Below that the asymptotic state of the second model is positive both in the exact and the weak coupling limit. 
 
For the presence of  $\gamma_\text{crit}$ in the considered model one can argue that for two oscillators, i.e., $n=1$ in Eq.  \eqref{eq:full_hamiltonian}, the coupling constant $C_1$ cannot be arbitrarily big. Above a well-defined value for $|C_1|$ the system is not fully oscillating: exponentially increasing and decreasing solutions appear among the particular solutions and this is true for the classical version of \eqref{eq:full_hamiltonian} with $n=1$, too. In this region of parameters, the system is unstable. This behavior is also expected for many bath oscillators: eventually the system is unstable when the couplings $C_n$ are increased beyond a certain value. This argument does not work for the second model. 
Here the singularities in $\Omega_\text{obs}^2(t)$  are due to the denominator of $A(t)$ in Eq. (\ref{eq:result_for_a(t)}), which becomes zero from time to time.  This, for example, cannot be explained by simple classical arguments. 

The validity of the third model in the parameter space was studied in detail by us \cite{HCsCsB}. At $T=0$ it was found that the allowed parameter range of $\gamma$ shrinks to the single point of $\gamma=0$. Here we have found that the first and second models have a finite range in $\gamma$ even at $T=0$. Surprisingly, the short-time behaviors of the first and the third models are the same; see Sec. \ref{subsec:short_time_expansions}.    

\section{Conclusions}
\label{V}

In the context of the quantum Brownian motion, we have studied the exact master equation of Hu, Paz, and Zhang for determining consistent ranges of the parameters of the model. By consistency we mean a time evolution, which maps density operators to density operators and the central system preserves its characteristics, i.e., a trapped particle remains trapped. The alternative approach of  Halliwell and Yu \cite{Halliwell2012} for determining the coefficients of the HPZ master equation offers succinct and tractable formulas. Therefore, we  started our paper with a brief review of their results. We  specified our analysis to a central harmonic oscillator and heat bath with a Lorentz-Drude type Ohmic spectral density \cite{Weiss,book1,HCsCsB} and $T=0$. Previous studies focused either on numerical evaluations of these coefficients \cite{HuPazZhang92} or analytical forms for simpler but unphysical spectral densities \cite{Ford}. The derivation of either an exact or approximated master equation was central to these studies \cite{book1}, but here, we lay emphasis on the consistency of the evolution itself. 

Based on this mindset, we have demonstrated that there exists a critical coupling constant $\gamma_\text{crit}$, above which the evolution of the system is nonphysical. These critical values persist for any finite temperature. The reason is that the observed  squared frequencies $\Omega_\text{obs}^2(t)$ (or its stationary value) are affected by the coefficient $A(t)$ only. The form of $A(t)$ depends on the temperature-independent kernel $\eta(s)$ directly or indirectly via the functions $u_1(s)$ and $u_2(s)$. Consequently, the nonphysical behavior in $\Omega_\text{obs}^2(t)$ persists at any temperature. Temperature-dependent effects can cause further nonphysical behaviors in any other physical quantities, but those effects can only narrow the allowed parameter regions. It turns out that not only the considered model in \eqref{eq:full_hamiltonian} but also the slightly changed Hamiltonian in \eqref{eq:second_hamiltonian}, i.e., the original Caldeira-Leggett model, has this critical value, too.

We have also shown how to calculate explicitly the four coefficients of the HPZ master equation for $T=0$. This analytical effort plays an important role in understanding the mathematical subtleties induced by different spectral densities, the transition from the exact to the weak coupling limit evolution, the short-time expansion of the coefficients, and the timescale under which the dynamics turn from a non-Markovian to a Markovian evolution. We have found that the short-time evolution is not affected by the weak coupling limit. However, $C(t)$, i.e., the cross-diffusion coefficient, and $D(t)$, i.e., the momentum diffusion coefficient, differ considerably over a long time from their weak-coupling counterparts. The other two coefficients A(t), i.e., the frequency shift, and $B(t)$, i.e., the time-dependent relaxation of the central system, are less influenced by the weak coupling approximation. We have also investigated two other versions of the model, where counterterms are added to the Hamiltonian of the central oscillator, and found that the weak coupling limit of one of these models can be extended naively beyond the allowed value of $\gamma_\text{crit}$. Usually, the four coefficients are investigated either numerically or established for easily tractable but nonphysical spectral densities, but in these approaches is hard to find the boundaries of the derived master equation. Here, the advantages of a complete analytical treatment are evident, though, we did not exploit its full potential. In general, questions related to consistency checks of master equations \cite{HCsCsB, BLH, Suarez, Pechukas, Hartmann, Jim_2022} should become more and more relevant due to a large amount of increasing activity in the area of non-Markovian evolution \cite{Rivas}, where various approximations are frequently used \cite{Vega, nat_commun}.

\begin{acknowledgments}
The authors are indebted to M. A. Csirik, M. Kornyik, and Z. Kaufmann for helpful discussions.
This research was supported by the National Research, Development and Innovation Office, Hungary (Project
No. TKP 2021-NVA-04) and the Frontline Research Excellence Programme of the NKFIH (Grant No.
KKP133827). A.C. acknowledges support from the National Research, Development and Innovation Office,
Hungary, Grant No. NKFI-134437. J.Z.B. acknowledges support from German Research Foundation under
Germany’s Excellence Strategy–Cluster of Excellence Matter and Light for Quantum Computing (ML4Q) EXC
2004/1-390534769, and AIDAS—AI, Data Analytics and Scalable Simulation—which is a Joint Virtual Laboratory
gathering the Forschungszentrum Jülich (FZJ) and the French Alternative Energies and Atomic Energy Commission (CEA).

\end{acknowledgments}

\appendix
\section{The structure of $G_1(s,\tau)$ and $G_2(s,\tau)$ for arbitrary spectral density}
\label{app:structures_of_G1_G2}

Let us use the notations of Sec. \ref{subsec:cal_of_u1_u2_g1_g2}. By Eqs. \eqref{eq:diff_eqs_for_u} and \eqref{eq:laplace_trafo} the Laplace transform of any solution $h(s)$ of the homogeneous equation \eqref{eq:diff_eqs_for_u} fulfills
\begin{equation}
    \left( z^2+\Omega^2+\frac{2}{M} \tilde{\eta}(z)\right) \tilde{h}(z)-h'(0)-z h(0)=0,
\end{equation}
where we have replaced the solution $u(s)$ of Eq. \eqref{eq:diff_eqs_for_u} by $h(s)$ for the general case. This implies that the function $h(s)$ defined by
\begin{equation}
    h(s)={\cal L}_z^{-1}\left[\tilde{h}(z)\right](s) \equiv {\cal L}_z^{-1}\left[ \frac{1}{z^2+\Omega^2+\frac{2}{M} \tilde{\eta}(z)}\right](s)
    \label{eq:def_of_h(s)}
\end{equation} 
has initial conditions
\begin{equation}
    h(0)=0, \quad h'(0)=1.
    \label{eq:init_conds_for_h}
\end{equation}
Taking the Laplace transform of Eq. \eqref{eq:diff_eqs_for_G},
\begin{eqnarray}
&&\left( z^2+\Omega^2+\frac{2}{M} \tilde{\eta}(z)\right) \tilde{G}_i(z,\tau)-G'_i(0,\tau)-z G_i(0,\tau)\nonumber \\
&&\quad =e^{-\tau z}\Theta (z),
\end{eqnarray}
$\tilde{G}_i(z,\tau)$ can be expressed as
\begin{eqnarray}
&&\tilde{G}_i(z,\tau)=\frac{e^{-\tau z}\Theta (z)+G'_i(0,\tau)+z G_i(0,\tau)}{z^2+\Omega^2+\frac{2}{M} \tilde{\eta}(z)} \nonumber \\
&&=\tilde{h}(z) \left( e^{-\tau z}\Theta (z)+G'_i(0,\tau)+z G_i(0,\tau)\right)
\label{eq:new_laplace_trafo_for_G_i}.
\end{eqnarray}
In the second equality we have used $\tilde{h}(z)$ defined by Eq. \eqref{eq:def_of_h(s)}.
Now, we use the translational property of the inverse Laplace transform for the first term: 
\begin{equation}
{\cal L}_z^{-1}\left[\tilde{h}(z)e^{-\tau z}\Theta(\tau)\right](s)=h(s-\tau)\Theta(\tau)\Theta(s-\tau).
\end{equation}
The identities ${\cal L}_s\left[h(s)\right](z)=\tilde{h}(z)$ and  ${\cal L}_s\left[h'(s)\right](z)=z\tilde{h}(z)+h(0)$ with $h(0)=0$ in Eq. \eqref{eq:init_conds_for_h} and
with the inverse Laplace transform of the second and third terms in \eqref{eq:new_laplace_trafo_for_G_i} yield
\begin{eqnarray}
&& {\cal L}_z^{-1}\left[\tilde{h}(z)\left(G'_i(0,\tau)+zG_i(0,\tau)\right)\right](s) \nonumber \\
&& \qquad=h(s)G'_i(0,\tau)+h'(s)G_i(0,\tau).
\end{eqnarray}
Regrouping the terms, we have:
\begin{eqnarray}
&&G_i(s,\tau)=h(s-\tau)\Theta(\tau)\Theta(s-\tau) \nonumber\\
&&\qquad +h(s)G'_i(0,\tau)+h'(s)G_i(0,\tau).
\label{eq:G_i_gen_case}
\end{eqnarray}
Taking derivative with respect to $s$, we obtain:
\begin{eqnarray}
&&G'_i(s,\tau)=h'(s-\tau)\Theta(\tau)\Theta(s-\tau)\nonumber \\
&&\qquad +h'(s)G'_i(0,\tau)+h''(s)G_i(0,\tau).
\label{eq:Gprime_i_gen_case}
\end{eqnarray}
The omitted term $h(s-\tau)\Theta(\tau)\delta(s-\tau)$ is zero due to \eqref{eq:init_conds_for_h}. The boundary conditions in \eqref{eq:bound_cond_for_G1} together with \eqref{eq:G_i_gen_case} result for $G_1(s,\tau)$:
\begin{equation}
    G_1(s,\tau)=\tilde{G}_1(s,\tau)\Theta(\tau)\Theta(s-\tau),
    \label{eq:G1_struct}
\end{equation}
with
\begin{equation}
    \tilde{G}_1(s,\tau)=h(s-\tau).
\end{equation}
Fixing the boundary conditions \eqref{eq:bound_cond_for_G2} for $G_2(s,\tau)$ at $s=t$, we require according to Eqs. \eqref{eq:G_i_gen_case} and \eqref{eq:Gprime_i_gen_case} that the currently unknown values $G_2(0,\tau)$  and $G_2'(0,\tau)$ are fixed by using the equations:
\begin{eqnarray}
&&0=G_2(t,\tau)=h(t-\tau)\Theta(\tau)\Theta(t-\tau) \nonumber\\
&&\qquad +h(t)G'_2(0,\tau)+h'(t)G_2(0,\tau), \\
&&0=G'_2(t,\tau)=h'(t-\tau)\Theta(\tau)\Theta(t-\tau) \nonumber\\
&&\qquad +h'(t)G'_2(0,\tau)+h''(t)G_2(0,\tau).
\end{eqnarray}
After solving $G_2(0,\tau)$ and $G_2'(0,\tau)$, we insert them into Eq. (\ref{eq:G_i_gen_case}) to have
\begin{equation}
    G_2(s,\tau)=G_1(s,\tau)+\tilde{G}_2(s,\tau) \Theta(\tau)\Theta(s-\tau),
\label{eq:G2_struct}
\end{equation}
with
\begin{eqnarray}
  && \tilde{G}_2(s,\tau) \nonumber \\
  &&=-\frac{h'(s)\begin{vmatrix} h(t-\tau) & h(t) \\ h'(t-\tau) & h'(t)\end{vmatrix} +h(s)\begin{vmatrix} h'(t) & h(t-\tau) \\ h''(t) & h'(t-\tau)\end{vmatrix}}{\begin{vmatrix} h'(t) & h(t) \\ h''(t) & h'(t)\end{vmatrix}}. \nonumber \\
\end{eqnarray}
Here $|\ldots|$ denotes the determinant of the matrices. Apart from the time windows the Heaviside functions
$G_1(s,\tau)$ and $G_2(s,\tau)$ are expressed in terms of $h(t)$, which is defined by the inverse Laplace transform in \eqref{eq:def_of_h(s)}; i.e., to calculate these Green's functions for a given spectral density $\tilde{\eta}(z)$ the inverse Laplace transform has to be performed.

\section{Careful evaluation of triple integrals}
\label{app:Careful_evaluation_triple_integrals}

The infinite upper bounds in single and triple integrals in Eq. \eqref{eq:def_of_c(t)} are moved down to some finite values due to the Heaviside functions in Eqs. \eqref{eq:g1_structure} and \eqref{eq:g2_structure}. Here we show the careful evaluation of the triple integrals by using only positive arguments of the kernel $\nu(s)$, where the necessary actual upper bounds are presented explicitly. Different upper bounds can be found in the literature (compare our formulas with \cite{Halliwell2012} and \cite{Fleming}, and also with the bounds in Ref. \cite{Chou2008}).
The same upper bounds must be used in triple integral for $D(t)$ with the trivial changes $G_1 \to G_1'$. The triple integral $J$ is
\begin{eqnarray}
    J&=&\int_0^t ds \int_0^{\infty} d\lambda\int_0^{\infty} d\tau \, \eta(t-s) \nonumber \\
    && \quad \times G_1(t,\lambda) G_2(s,\tau) \nu(\tau-\lambda),
\end{eqnarray}
where the kernel $\nu$ is symmetric, i.e.,
$\nu(\tau-\lambda)=\nu(\lambda-\tau)$.
The structures of $G_1$ and $G_2$ are given in Eqs. (\ref{eq:G1_struct}) and (\ref{eq:G2_struct}).
The original integral $J$ is obtained from these two terms as 
\begin{equation}
    J=J_1+J_2.
\end{equation}
$J_1$ contains the contribution of the first term in (\ref{eq:G2_struct}) and $J_2$ belongs to the second group. 
The necessary integrals using only the smooth functions $\tilde{G}_1(s,\tau)$ and $\tilde{G}_2(s,\tau)$ are
\begin{eqnarray}
&&J_1= \nonumber \\
&&\int_0^t ds \int_0^s d\tau \int_0^{\tau} d\lambda \, \eta(t-s)\tilde{G}_1(t,\lambda)\tilde{G}_1(s,\tau)\nu(\tau-\lambda) \nonumber \\
&&+ \int_0^t ds \int_0^s d\tau \int_{\tau}^t d\lambda \, \eta(t-s)\tilde{G}_1(t,\lambda)\tilde{G}_1(s,\tau)\nu(\lambda-\tau),\nonumber \\
\label{eq:J_1_app}
\end{eqnarray}
and
\begin{eqnarray}
&&J_2=\nonumber \\
&&\int_0^t ds \int_0^t d\tau \int_0^{\tau} d\lambda \, \eta(t-s)\tilde{G}_1(t,\lambda)\tilde{G}_2(s,\tau)\nu(\tau-\lambda) \nonumber \\
&&+ \int_0^t ds \int_0^t d\tau \int_{\tau}^t d\lambda \, \eta(t-s)\tilde{G}_1(t,\lambda)\tilde{G}_2(s,\tau)\nu(\lambda-\tau).\nonumber \\
\label{eq:J_2_app}
\end{eqnarray}
Now, the bounds of the integrals ensure causality; we have no contributions to the master equation's coefficients from later times than $t$.   

\section{Some explicit intermediate coefficients}
\label{app:exp_ceffs}

Coefficients appearing in Eqs. \eqref{eq:C_op_act_to_exponential} and \eqref{eq:D_op_act_to_exponential} read
\begin{eqnarray}
C_{1,1}(\alpha)&=&\frac{\Omega_c+z_1}{(\alpha-z_1)(z_1-z_2)(z_3-z_1)}, \\
D_{1,1}(\alpha)&=&\frac{(\Omega_c+z_1)z_1}{(\alpha-z_1)(z_1-z_2)(z_3-z_1)}.
\end{eqnarray}
The functions $C_{1,2}(\alpha)$ and $D_{1,2}(\alpha)$ are obtained from $C_{1,1}(\alpha)$ and $D_{1,1}(\alpha)$ by applying the simultaneous replacements $(z_1 \to z_2,z_2 \to z_3,z_3 \to z_1)$. In a similar way, $C_{1,3}(\alpha)$ and $D_{1,3}(\alpha)$ are the results of the simultaneous replacements $(z_1 \to z_3,z_2 \to z_1,z_3 \to z_2)$.
The remaining coefficients $C_{1,0}(\alpha)$ and $D_{1,0}(\alpha)$ are given by
\begin{eqnarray}
C_{1,0}(\alpha)&=&-\Bigl(C_{1,1}(\alpha)+C_{1,2}(\alpha)+C_{1,3}(\alpha)\Bigr), \\
D_{1,0}(\alpha)&=&-\Bigl(D_{1,1}(\alpha)+D_{1,2}(\alpha)+D_{1,3}(\alpha)\Bigr).
\end{eqnarray}
Coefficients appearing on the right-hand sides of Eqs. \eqref{eq:C_3_op_act_to_exponential} and \eqref{eq:D_3_op_act_to_exponential}
are
\begin{widetext}
\begin{equation}
C_{3,1}(\alpha)=-M \gamma \Omega_c^2 \frac{ (z_1-z_2)(z_1-z_3)(z_2-z_3)^2 [\alpha^2+(z_1+z_2+z_3)\alpha+(z_1+z_2)(z_1+z_3)]   (z_1+\Omega_c)}{z_1(z_1+z_2) (z_1+z_3) (z_1+\alpha) (z_2+\alpha) (z_3+\alpha)}, 
\end{equation}
\begin{equation}
\quad D_{3,1}(\alpha)=z_1 C_{3,1}(\alpha),
\end{equation}
\begin{equation}
C_{3,4}(\alpha)=-M \gamma \Omega_c^2 \frac{ (z_1-z_2)^2(z_1-z_3)(z_2-z_3) [(\Omega_c-z_3) \alpha^2+(z_1+z_2+z_3)(z_3-\Omega_c)\alpha+(z_1+z_3)(z_2+z_3)\Omega_c]}{z_3(z_1+z_3) (z_2+z_3) (z_1-\alpha) (z_2-\alpha) (z_3-\alpha)}, 
\end{equation}
\begin{equation}
\quad D_{3,4}(\alpha)=-M \gamma \Omega_c^2 \frac{ (z_1-z_2)^2(z_1-z_3)(z_2-z_3)\alpha [(z_3-\Omega_c)\alpha+z_1 z_2+(z_1+z_2+z_3)\Omega_c ]}{(z_1+z_3) (z_2+z_3) (z_1-\alpha) (z_2-\alpha) (z_3-\alpha)},
\end{equation}
\begin{equation}
C_{3,7}(\alpha)=M\gamma\Omega_c^2\frac{(z_1-z_2)(z_2-z_3)(z_1+\Omega_c)}{(z_1-\alpha)(z_3+\alpha)},
\end{equation}
\begin{equation}
D_{3,7}(\alpha)=z_1 C_{3,7}(\alpha),
\end{equation}
\begin{equation}
C_{3,10}(\alpha)=M\gamma\Omega_c^2\frac{(z_1-z_3)(z_3-z_2)(z_1+\Omega_c)}{(z_1-\alpha)(z_2+\alpha)},
\end{equation}
\begin{equation}
D_{3,10}(\alpha)=z_1 C_{3,10}(\alpha).
\end{equation}

The coefficients $C_{3,i}$ and $D_{3,i}$ for $i=2$, $i=5$, $i=8$, and $i=11$ are obtained from the coefficients of $i=1$, $i=4$, $i=7$, and $i=10$   
by applying the simultaneous replacements $(z_1 \to z_2,z_2 \to z_3,z_3 \to z_1)$, respectively. Similarly, coefficients for $i=3$, $i=6$, $i=9$, and  $i=12$ are obtained from the coefficients of $i=1$, $i=4$, $i=7$, and $i=10$   
by applying the simultaneous replacements $(z_1 \to z_3,z_2 \to z_1,z_3 \to z_2)$. The most complex expressions are the closed forms of $C_{3,13}(\alpha)$ and $D_{3,13}(\alpha)$:
\begin{equation}
C_{3,13}(\alpha)=M\gamma\Omega_c^2 \left[ \left(\frac{(z_2 - z_3)^2 (z_1^2 + z_2 z_3) (z_1 + \Omega_c)}{z_1 (z_1 + z_2) (z_1 + z_3) (z_1 + \alpha)} + \frac{(z_2 - z_3)^2 (z_2 z_3 \Omega_c + z_1^2 (z_2 + z_3 + \Omega_c))}{z_1 (z_1 + z_2) (z_1 + z_3) (z_1  -\alpha)}\right)+\textrm{Cycl.}\right],
\end{equation}
\begin{equation}
D_{3,13}(\alpha)=M\gamma\Omega_c^2 \left[ \left(\frac{(z_2 - z_3)^2 (z_1^2 + z_2 z_3) (z_1 + \Omega_c)}{(z_1 + z_2) (z_1 + z_3) (z_1 + \alpha)} + \frac{z_1(z_2 - z_3)^2 (z_1^2+z_2 z_3+(z_2+z_3)\Omega_c)}{ (z_1 + z_2) (z_1 + z_3) (z_1  -\alpha)}\right)+\textrm{Cycl.}\right].
\end{equation}
\end{widetext}

\section{Useful integrals}
\label{app:useful_integrals}

Let us suppose that $r$ is a complex parameter and $t$ is positive. We need some expression for the integrals
\begin{equation}
I_1(r,t)=\int_0^\infty d\omega\, \frac{\omega \cos(\omega t)}{r^2+\omega^2}, \qquad t>0 , \quad r \in \mathbb{C}
\label{eq:def_of_i1}
\end{equation}
and
\begin{equation}
I_2(r,t)=\int_0^\infty d\omega\, \frac{\omega \sin(\omega t)}{r^2+\omega^2}, \qquad t>0 , \quad r \in \mathbb{C}.
\label{eq:def_of_i2}
\end{equation}
Both integrals are invariant under the change $r \to -r$, thus it is enough to consider the integrals in the region $-\pi/2 < \arg r <\pi/2$, which we consider throughout this Appendix. By writing the fraction as a sum of partial fractions one can shift the integration variables so that the integrals can be written in terms of the functions
\begin{equation}
\mathrm{Chi}\,(z)=-\gamma_\text{EM} +\ln(z)+\int_0^z \frac{\cosh(t)-1}{t}dt,
\label{eq:def_chi}
\end{equation}
called hyperbolic cosine integral and
\begin{equation}
\mathrm{Shi}\,(z)=\int_0^z \frac{\sinh(t)}{t}dt,
\label{eq:def_shi}
\end{equation}
called hyperbolic sine integral. Useful formulas especially their behaviors at infinity can be found, e.g., in Ref. \cite{wolfram}. The integrals in question are expressed as
\begin{equation}
I_1(r,t)=-\left[\cosh(rt)\mathrm{Chi}\,(rt)-\sinh(rt)\mathrm{Shi}\,(rt) \right],
\label{eq:i1_expr}
\end{equation}
\begin{equation}
I_2(r,t)=-\frac{1}{r}\left[\sinh(rt)\mathrm{Chi}\,(rt)-\cosh(rt)\mathrm{Shi}\,(rt) \right].
\label{eq:i2_expr}
\end{equation}
In analytical calculations for big enough $t$ one faces the problem that on the right-hand sides of Eqs. \eqref{eq:i1_expr} and \eqref{eq:i2_expr} one has the difference of two very big quantities; however, the difference is small in the right half plane for the complex $r$. One can find useful asymptotic expansions for both integrals using, e.g.,  Mathematica software \cite{wolfram}.


\begin{thebibliography}{100}
\bibitem{Weiss} U. Weiss, {\it Quantum Dissipative Systems}, 2nd ed. (World Scientific, Singapore, 1999).

\bibitem{book1} H.-P. Breuer and F.~Petruccione, {\it The Theory of Open Quantum Systems} (Oxford University Press, Oxford, 2002).

\bibitem{Grabert} H. Grabert, P. Schramm, and G.-L. Ingold, Phys. Rep. {\bf 168}, 115 (1988).

\bibitem{CL} A. O. Caldeira and A. J. Leggett, Physica {\bf 121A}, 587 (1983).

\bibitem{HuPazZhang92} B. L. Hu, J. P. Paz, and Y. Zhang, Phys. Rev. D {\bf 45}, 2843 (1992).

\bibitem{Halliwell2012} J. J. Halliwell and T. Yu, Phys. Rev. D {\bf 53}, 2012 (1996).

\bibitem{Anglin} J. Anglin and S. Habib, Mod. Phys. Lett. A {\bf 11}, 2655 (1996).

\bibitem{Fleming} C. H. Fleming, A. Roura, and B. L. Hu, Ann. Phys. {\bf 326}, 1207 (2011).

\bibitem{Nakajima} S. Nakajima, Progr. Theor. Phys. {\bf 20}, 948 (1958). 

\bibitem{Zwanzig} R. Zwanzig, J. Chem. Phys. {\bf 33}, 1338 (1960).

\bibitem{Prigogine} I. Prigogine, {\it Non-Equilibrium Statistical Mechanics} (Interscience Publishers, New York, 1962).

\bibitem{Anders} J. Anders, C. R. J. Sait, and S. A. R. Horsley,  New J. Phys.~{\bf 24}, 033020 (2020).

\bibitem{Gottwald}  F. Gottwald, S. D. Ivanov, and O. Kühn, J. Phys. Chem. Lett.~{\bf 6}, 2722 (2015).

\bibitem{Rognoni} A. Rognoni, R. Conte, and M. Ceotto, J. Chem. Phys.~{\bf 154}, 094106 (2021).

\bibitem{Bothma}  J. P. Bothma, J. B. Gilmore, and R. H. McKenzie, New J. Phys.~{\bf 12}, 055002 (2010).

\bibitem{Hanggi} P. H\"anggi, P. Talkner, and M. Borkovec, Rev. Mod. Phys.~{ \bf 62}, 251 (1990).

\bibitem{Giesel} K. Giesel and M. Kobler, Mathematics~{\bf 10}, 4248 (2022).

\bibitem{Jahn} M. J. Fahn, K. Giesel and M. Kobler, Class. Quantum Grav.~{ \bf 40}, 094002 (2023).

\bibitem{Yabu} H. Yabu, K. Nozawa, T. Suzuki, Phys. Rev. D~{\bf 57}, 1687 (1998).

\bibitem{FordKac} G. W. Ford, M. Kac, and P. Mazur, J. Math. Phys. {\bf 6}, 504 (1965).

\bibitem{Ullersma} P. Ullersma, Physica {\bf 32}, 27 (1966); {\bf 32}, 56 (1966); {\bf 32}, 74 (1966); {\bf 32}, 90 (1966).

\bibitem{Ford01} G. W. Ford and M. Kac, J. Stat. Phys. {\bf 46}, 803 (1987).

\bibitem{Kappler} H.-P. Breuer, B. Kappler, and F. Petruccione, Ann. Phys.  (N.Y.) {\bf 291}, 36 (2001).

\bibitem{Ford} G. W. Ford and R. F. O'Connell, Phys. Rev. D {\bf 64}, 105020 (2001).

\bibitem{HCsCsB} G. Homa, A. Csord\'as, M. A. Csirik, and J. Z. Bern\'ad, Phys. Rev. A {\bf 102}, 022206 (2020).

\bibitem{Eisert} S. Gr\"oblacher, A. Trubarov, N. Prigge, G. D. Cole, M. Aspelmeyer, and J. Eisert, Nat. Commun. {\bf 6}, 7606 (2015).

\bibitem{Paris} M. Bina, F. Grasselli, and M. G. A. Paris, Phys. Rev. A {\bf 97}, 012125 (2018).

\bibitem{Lampo} A. Lampo, M. Garcia-March, and M. Lewenstein, {\it Quantum Brownian Motion Revisited: Extensions and Applications} (Springer, Cham, 2019).

\bibitem{Lindblad} G. Lindblad, Comm. Math. Phys. {\bf 48}, 119 (1976).

\bibitem{Evans} D. E. Evans, Acta. Appl. Math {\bf 2}, 333 (1984).

\bibitem{Walls} D. F. Walls and G. J. Milburn, Phys. Rev. A {\bf 31}, 2403 (1985).

\bibitem{Ford02} G. W. Ford and R. F. O'Connell, J. Optics B. {\bf 5}, 609 (2003).


\bibitem{Courant} R. Courant and D. Hilbert, {\it Methods of Mathematical Physics}
(Interscience Publishers, New York, 1962).

\bibitem{HCSB} J. Z. Bern\'ad, G. Homa, and M. A. Csirik, Eur. Phys. J. D {\bf 72}, 212 (2018).



\bibitem{Newton} G. Homa {\it et al}., J. Phys. A: Math. Theor. {\bf 56}, 145203 (2023).

\bibitem{caruso2018} F. Caruso, J. Eisert, V. Giovannetti, and A.S. Holevo, New J. Phys. {\bf10}, 083030, (2018).

\bibitem{Halliwellarxiv} J. J. Halliwell and T. Yu, arXiv:quant-ph/9508004.

\bibitem{Davies} B. Davies, {\it Integral Transforms and Their Applications} (Springer, Berlin, 1978).

\bibitem{Unruh} W. G. Unruh and W. H. Zurek, Phys. Rev. D {\bf 40}, 1071 (1989).

\bibitem{BLH} G. Homa, J. Z. Bern\'ad and L. Lisztes, Eur. Phys. J. D {\bf 73}, 53 (2019).

\bibitem{Suarez} A. Su\'arez, R. Silbey, and I. Oppenheim, J. Chem. Phys. {\bf 97}, 5101 (1992).

\bibitem{Pechukas} P. Pechukas, Phys. Rev. Lett. 
{\bf 73}, 1060 (1994).

\bibitem{Hartmann} R. Hartmann and W. T. Strunz, 
Phys. Rev. A {\bf 101}, 012103 (2020).

\bibitem {Jim_2022} S. Lally, N. Werren, J. Al-Khalili, and A. Rocco, Phys. Rev. A {\bf 105}, 012209 (2022).

\bibitem{Rivas} A. Rivas, S. F. Huelga, and M. B. Plenio, Phys. Rev. Lett. {\bf 105}, 050403 (2010).

\bibitem{Vega} I. de Vega and D. Alonso, Rev. Mod. Phys. {\bf 89}, 015001 (2017).

\bibitem{nat_commun} A. Strathearn, P. Kirton, D. Kilda, J. Keeling and  B. W. Lovett, Nat. Commun. {\bf 105},  3322 (2018).

\bibitem{Chou2008} C.-H. Chou, T. Yu, and B. L. Hu, Phys. Rev. E {\bf 77}, 011112 (2008). 

\bibitem{wolfram} http://mathworld.wolfram.com.

\end{thebibliography}
\end{document}